\documentclass[12pt]{article}
\usepackage{ctex}
\usepackage{amssymb,amsmath,color,amsmath,bm}
\usepackage[colorlinks,linkcolor=blue]{hyperref}
\usepackage{geometry}
\usepackage{tabularx}
\usepackage{bm}
\usepackage{authblk}
\usepackage[square, comma, sort&compress, numbers]{natbib}

\usepackage{float}  

	\newcommand{\rank}{{\mathrm{rank}}}
	\newtheorem{theorem}{Theorem}[section]
	\newtheorem{definition}{Definition}[section]
	\newtheorem{lemma}{Lemma}[section]
	\newtheorem{example}{Example}[section]
	
	\newtheorem{corollary}{Corollary}[section]
	\newtheorem{remark}{Remark}[section]

	\catcode`@=11 \@addtoreset{equation}{section} \catcode`@=12
\begin{document}
\title{\bf The extended code for a class of generalized Roth-Lempel codes and their properties\thanks{This paper is supported by National Natural Science Foundation of China (Grant No. 12471494) and Natural Science Foundation of Sichuan Province (2024NSFSC2051). The corresponding author is Professor Qunying Liao.}}
\author{\small Zhonghao Liang}
\author{\small Qunying Liao
{\thanks{Z. Liang and Q. Liao are with the College of Mathematical Science, Sichuan Normal University, Chengdu 610066, China (e-mail:liangzhongh0807@163.com;qunyingliao@sicnu.edu.cn)}}
}
\affil[] {\small(College of Mathematical Sciences, Sichuan Normal University, Chengdu, 610066, China)}
\date{}
\maketitle
	{\bf Abstract.}
	{\small As we all know, many interesting and important codes are obtained by modifying or combining existing codes. In this paper, we focus on generalized Roth-Lempel (in short, GRL) codes and define a class of extended codes, i.e., the extended generalized Roth-Lempel (in short, EGRL) code. And then for a  special class of EGRL codes, we give a  parity-check matrix and establish a necessary and sufficient condition for the EGRL code or its dual code to be MDS or AMDS, respectively. Finally, we construct a class of NMDS EGRL codes which is the generalization of the constructions given by Han et al. in 2023, and then completely determine its weight distribution.   
 }\\
	

\section{Introduction}
Let $\mathbb{F}_q$ be the finite field with $q=p^s$, where $p$ is a prime and $s$ is a positive integer. Let $\mathbb{F}_{q}^{*}=\mathbb{F}_{q}\backslash\left\{0\right\}$. A non-empty set $\mathcal{C}$ of $\mathbb{F}_q$ is called an $[n,k,d]$ linear code over $\mathbb{F}_q^n$ if  $\mathcal{C}$ is a $k$-dimensional linear subspace over $\mathbb{F}_q^n$, where $k$ and
$d$ are referred to as the dimension and the minimum distance of $\mathcal{C}$, respectively. The dual code of $\mathcal{C}$
is defined as
\[
\mathcal{C}^{\perp}=\left\{\left(x_{1}, \ldots, x_{n}\right)=\boldsymbol{x}\in\mathbb{F}_q^{n} \mid\langle\boldsymbol{x},\boldsymbol{y}\rangle=\sum\limits_{i=1}^{n} x_{i} y_{i}=0,  \forall \boldsymbol{y}=\left(y_{1}, \ldots, y_{n}\right) \in \mathcal{C}\right\}.
\]

For a codeword $\boldsymbol{c}=(c_1,\ldots, c_n) \in \mathcal{C}$, its support is defined as $\operatorname{supp}(\boldsymbol{c}) = \{1 \leq i \leq n|c_i \neq 0\}$, $\operatorname{wt}(\boldsymbol{c}) = \#\operatorname{supp}(\boldsymbol{c})$ is the Hamming weight of $\boldsymbol{c}$ and the minimum Hamming distance $d(\mathcal{C})$ is the minimal Hamming weight of all non-zero codewords of  $\mathcal{C}$. Let $A_i(1\leq i\leq n)$ be the number of codewords with Hamming weight $i$ in $\mathcal{C}$. Then the weight distribution of $\mathcal{C}$ is characterized by the sequence $(1,A_1,\ldots, A_n)$ and the polynomial $ A(x) = 1 + A_1x + A_2x^2 + \cdots + A_nx^n$ is called the weight enumerator of $\mathcal{C}$.

For an $[n, k, d]$  linear code $\mathcal{C}$, the Singleton bound states that  $d\leq n-k+1$. If $d=n-k+1$, i.e., the Singleton bound is reached, then $\mathcal{C}$ is maximum distance separable (in short, MDS). If  $d=n-k$, then $\mathcal{C}$ is almost MDS (in short, AMDS). Especially, if $\mathcal{C}^{\perp}$ is also AMDS, then $\mathcal{C}$ is near MDS (in short, NMDS). 

Since both MDS codes and NMDS codes have good properties, they are important in coding theory
and the other related fields \cite{A1,A2}, and so their constructions are interesting  \cite{A3,A4,A5,A6,A7,A8,A9,A10,A11,A12,A13,A14,A15,A16}. Especially in recent years, to construct MDS or NMDS codes based on generalized Reed-Solomon codes (in short, GRS) or extended generalized Reed-Solomon codes
(in short, EGRS), have attracted much attention  \cite{A17,A18,A19,A20,A21,A22,A23,A24,A25}. Additionally, basing on the generator matrix of RS codes,  in 1989, Roth and Lempel \cite{A26} introduced a new construction of
MDS codes by adding two columns into the generator matrix
of RS codes, i.e., the corresponding linear code over $\mathbb{F}_{q}$ has the generator matrix
$$
\begin{pmatrix}\begin{matrix}
	\boldsymbol{G}_{RS}(\boldsymbol{\alpha})
\end{matrix}&\begin{matrix}
\boldsymbol{0}_{(k-2)\times 2}\\
\boldsymbol{T}_{2\times 2}(\delta)
\end{matrix}
\end{pmatrix}_{k\times (n+2)},
$$
\iffalse
\begin{equation}\label{RS}
\begin{pmatrix}
		1&	\cdots&		1&0&0\\
		\alpha_1&	\cdots&\alpha _n&0&0\\
		\vdots&		\quad&		\vdots&		\vdots&		\vdots\\
		\alpha _{1}^{k-3}&	\cdots&\alpha _{n}^{k-3}&0&0\\
		\alpha _{1}^{k-2}&	\cdots&\alpha _{n}^{k-2}&0&1\\
		\alpha _{1}^{k-1}&	\cdots&\alpha _{n}^{k-1}&1&\delta\\
	\end{pmatrix}_{k\times n},
\end{equation}\fi
and is denoted by $\mathrm{RL}_{k}(\boldsymbol{\alpha},\delta)$, where $\boldsymbol{G}_{RS}(\boldsymbol{\alpha})$ is the generator matrix of RS codes with the evaluation-point sequence $\boldsymbol{\alpha}=\left(\alpha_{1}, \ldots, \alpha_{n}\right) \in \mathbb{F}_{q}^{n}$, $\boldsymbol{T}_{2\times 2}(\delta)=\begin{pmatrix}
	0&1\\
	1&\delta
\end{pmatrix}, \delta \in\mathbb{F}_{q}$. At the same time, they also gave an equivalent condition for $\mathrm{RL}_{k}(\boldsymbol{\alpha},\delta)$ to be MDS. In 2023, by properties of the projective space, Han et al. \cite{A4} presented a necessary and 
sufficient condition for $\mathrm{RL}_{k}(\boldsymbol{\alpha},0)$ to be NMDS, constructed a class of NMDS Roth-Lempel codes $\mathrm{RL}_{k}(\mathbb{F}_{q},0)$ and completely determined its weight distribution, the corresponding linear code over $\mathbb{F}_{q}$ has the generator matrix
\begin{equation}\label{2023HanRLcode}
	\begin{pmatrix}\begin{matrix}
			\boldsymbol{G}_{RS}(\mathbb{F}_{q})
		\end{matrix}&\begin{matrix}
			\boldsymbol{0}_{(k-2)\times 2}\\
			\boldsymbol{T}_{2\times 2}(0)
		\end{matrix}
	\end{pmatrix}_{k\times (q+2)},
\end{equation}
\iffalse$$\begin{pmatrix}
	1 & \cdots & 1&0&0\\
	\beta_{1} & \cdots & \beta_{q}&0&0 \\ 
	\vdots & \ddots & \vdots & \vdots& \vdots\\
	
	\beta_{1}^{k-4} & \cdots & \beta_{q}^{k-4}&0&0\\
	\beta_{1}^{k-3} & \cdots & \beta_{q}^{k-3}&0&0\\
	\beta_{1}^{k-2} & \ldots & \beta_{q}^{k-2}&0&1\\
	\beta_{1}^{k-1} & \cdots & \beta_{q}^{k-1}&1&0
	
\end{pmatrix}_{k\times (q+2)}.$$\fi
 where $\mathbb{F}_{q}=\left\{\beta_{1},\ldots,\beta_{q}\right\}$. Especially, for the case $\beta_{q}=0$, Zhang et al. \cite{A10} proved the same result by the definition of the linear code. In 2024, Wu et al. \cite{A6} constructed a class of the extended code of $\mathrm{RL}_{k}(\boldsymbol{\alpha},\delta)$, the corresponding linear code over $\mathbb{F}_{q}$ has the generator matrix 
\begin{equation}\label{2024WURLcode}
	\begin{pmatrix}\begin{matrix}
			\boldsymbol{G}_{RS}(\boldsymbol{\alpha})
		\end{matrix}&\begin{matrix}
			\boldsymbol{0}_{(k-3)\times 3}\\
			\boldsymbol{T}_{3\times 3}(\delta,\tau,\mu)
		\end{matrix}
	\end{pmatrix}_{k\times (n+3)},
\end{equation}
\iffalse\begin{equation}\label{WuDingmatrix}
\begin{pmatrix}
	1 & \cdots & 1&0&0&0\\
	\alpha_{1} & \cdots & \alpha_{n}&0&0&0 \\ 
	\vdots & \ddots & \vdots& \vdots & \vdots& \vdots\\		
	\alpha_{1}^{k-4} & \cdots & \alpha_{n}^{k-4}&0&0&0\\
	\alpha_{1}^{k-3} & \cdots & \alpha_{n}^{k-3}&0&0&1\\
	\alpha_{1}^{k-2} & \ldots & \alpha_{n}^{k-2}&0&1&\tau\\
	\alpha_{1}^{k-1} & \cdots & \alpha_{n}^{k-1}&1&\delta&\mu
\end{pmatrix}_{k\times (n+3)},
\end{equation}\fi
where $\boldsymbol{T}_{3\times 3}(\delta,\tau,\mu)=\begin{pmatrix}
	0&0&1\\
	0&1&\tau\\
	1&\delta&\mu
\end{pmatrix}, \delta,\tau,\mu\in\mathbb{F}_{q}$. And then they also gave an equivalent condition for the corresponding linear code or its dual code to be MDS or AMDS, respectively. Recently, the authors introduced the definition of the generalized Roth-Lempel
 (in short, GRL) code in \cite{A16} and constructed two classes of NMDS GRL codes in \cite{A15}.
   
In addition, an important topic in coding theory is to construct linear codes with
interesting parameters and properties over $\mathbb{F}_{q}$. There are many ways of constructing good linear codes based on a given linear code $\mathcal{C}$ over $\mathbb{F}_{q}$, such as by puncturing  \cite{A27,A28}, or shortening \cite{A29,A30,A31}, or extending the given linear code $\mathcal{C}$  \cite{A32,A33,A34} and so on. 

Motivated by the above works, we introduce the concept of the extended generalized Roth-Lempel
(in short, EGRL) code. And then for a special class of EGRL codes, we give an equivalent condition for the EGRL code to be MDS or AMDS, respectively. Finally, we construct a class of NMDS EGRL codes and completely determine its weight distribution.

This paper is organized as follows. In Section 2, we give the definition of the extended generalized Roth-Lemmpel code and some necessary lemmas. In Sections 3-4, we give a sufficient and necessary condition for $\mathrm{EGRL}_{k,2,0}\left(\boldsymbol{\alpha},\boldsymbol{1},b\right)$ or its dual code to be MDS or AMDS, respectively.
In Section 5, we construct a class of NMDS EGRL codes $\mathrm{EGRL}_{k,2,0}\left(\mathbb{F}_{q}^{*},\boldsymbol{1},b\right)$ and determine its weight distribution. In Section 6, we conclude the whole paper.
	\section{Preliminaries}
In this section, we give the definition of the EGRL code and some necessary lemmas.
\begin{definition}\label{EGRLdefinition}
	Let $\mathbb{F}_q$ be the finite field of $q$ elements, where $q$ is a prime power. Let $l+1\leq k+1\leq n\leq q$,  $\boldsymbol{\alpha}=\left(\alpha_{1}, \ldots, \alpha_{n}\right) \in \mathbb{F}_{q}^{n}$ with $\alpha_{i} \neq \alpha_{j}(i \neq j)$, $b\in\mathbb{F}_{q}^{*}$, $\boldsymbol{v}=$ $\left(v_{1}, \ldots, v_{n}\right) \in\left(\mathbb{F}_{q}^{*}\right)^{n}$ and $\boldsymbol{M}_{l\times l}=(a_{ij})_{l\times l}\in\mathrm{GL}_{l}\left(\mathbb{F}_{q}\right)$. The  extended generalized Roth-Lempel (in short, EGRL) code $\mathrm{EGRL}_{k,\ell,t}(\boldsymbol{\alpha}, \boldsymbol{v},b)$ is defined as 
	$$
	\mathrm{EGRL}_{k,\ell,t}(\boldsymbol{\alpha}, \boldsymbol{v},b)\triangleq \left\{\left(v_{1} f\left(\alpha_{1}\right), \ldots, v_{n} f\left(\alpha_{n}\right),\boldsymbol{\beta},bf_{t}\right) | f(x)\in \mathbb{F}_{q}^{k}[x]\right\},
	$$
where $0\leq t\leq k-3$, $\mathbb{F}_{q}[x]$ is the polynomial ring over $\mathbb{F}_{q}$,
$$\mathbb{F}_{q}^{k}[x]=\left\{f(x)=\sum\limits_{i=0}^{k-1} f_{i} x^{i}\mid f_{i} \in \mathbb{F}_{q}, 0 \leq i \leq k-1\right\}$$
and $$\begin{aligned}
	\boldsymbol{\beta}=&\left(f_{k-l},\ldots,f_{k-1}\right)\boldsymbol{M}_{l\times l}\\
	=&\left(a_{11}f_{k-l}+a_{21}f_{k-(l-1)}+\cdots+a_{l1}f_{k-1},\ldots,a_{1l}f_{k-l}+a_{2l}f_{k-(l-1)}+\cdots+a_{ll}f_{k-1}\right).
	\end{aligned}$$
\end{definition} 
\begin{remark}\label{generatormatrix}
$(1)$ By taking $(\ell,b,t)=(2,1,k-3)$, $\boldsymbol{v}=\boldsymbol{1}\in\mathbb{F}_{q}^{n}$ and $\boldsymbol{M}_{l\times l}= \begin{pmatrix}
	0&1\\
	1&\delta
\end{pmatrix}$ in Definition $\ref*{EGRLdefinition}$, the corresponding code $\mathrm{EGRL}_{k,2,k-3}(\boldsymbol{\alpha}, \boldsymbol{1},1)$ has the generator matrix
$$
	\begin{pmatrix}
		1 & \cdots & 1&0&0&0\\
		\alpha_{1} & \cdots & \alpha_{n}&0&0&0 \\ 
		\vdots & \ddots & \vdots& \vdots & \vdots& \vdots\\
		
		\alpha_{1}^{k-4} & \cdots & \alpha_{n}^{k-4}&0&0&0\\
		\alpha_{1}^{k-3} & \cdots & \alpha_{n}^{k-3}&0&0&1\\
		\alpha_{1}^{k-2} & \ldots & \alpha_{n}^{k-2}&0&1&0\\
		\alpha_{1}^{k-1} & \cdots & \alpha_{n}^{k-1}&1&\delta&0
		
	\end{pmatrix}_{k\times (n+3)},
$$
which is just a special case of the matrix given by $(\ref{2024WURLcode})$, i.e., $\tau=\mu=0$.

$(2)$ By taking $(\ell,b,t)=(2,1,0)$,  $\boldsymbol{\alpha}=\left(\alpha_{1}, \ldots, \alpha_{n}\right)$ with $\left\{\alpha_{1},\ldots,\alpha_{n}\right\}=\mathbb{F}_{q}^{*},\boldsymbol{v}=\boldsymbol{1}\in\mathbb{F}_{q}^{n},\boldsymbol{M}_{l\times l}= \begin{pmatrix}
0&1\\
1&0
\end{pmatrix}$ in Definition $\ref*{EGRLdefinition}$, the corresponding code $\mathrm{EGRL}_{k,2,0}(\mathbb{F}_{q}^{*}, \boldsymbol{1},1)$ has the generator matrix 
$$
\begin{pmatrix}
	1 & \cdots & 1&0&0&1\\
	\alpha_{1} & \cdots & \alpha_{q-1}&0&0&0 \\ 
	\vdots & \ddots & \vdots& \vdots & \vdots& \vdots\\
	
	\alpha_{1}^{k-4} & \cdots & \alpha_{q-1}^{k-4}&0&0&0\\
	\alpha_{1}^{k-3} & \cdots & \alpha_{q-1}^{k-3}&0&0&0\\
	\alpha_{1}^{k-2} & \ldots & \alpha_{q-1}^{k-2}&0&1&0\\
	\alpha_{1}^{k-1} & \cdots & \alpha_{q-1}^{k-1}&1&0&0
	
\end{pmatrix}_{k\times (q+2)},$$
and is monomially equivalent to the linear code generated by the matrix in  $(\ref{2023HanRLcode})$.
\end{remark}
\begin{lemma}\label{monomiallyequivalent}
{\rm(\cite{A6}, Definition 3)} Let $\mathcal{C}_{1}$ and $\mathcal{C}_{2}$ be two linear codes  of the same length over $\mathbb{F}_{q}$, and let $\boldsymbol{M}$ be a generator matrix of $\mathcal{C}_{1}$. Then $\mathcal{C}_{1}$ and $\mathcal{C}_{2}$ are monomially equivalent if and only if there exists a monomial matrix $\boldsymbol{D}$ such that $\boldsymbol{MD}$ is a generator matrix of $\mathcal{C}_{2}$.
\end{lemma}

\begin{lemma}\label{MDSdefinition}
	{\rm(\cite{A1}, Theorem 2.4.3)}
Let $\mathcal{C}$ be an $[n, k]$ code over $\mathbb{F}_{q}$ with $k\geq 1$. Suppose that  $\boldsymbol{G}$  and  $\boldsymbol{H}$  are the 
generator matrix and parity-check matrix for $\mathcal{C}$, respectively. Then, the following statements are equivalent to each other,

$(1)\ \mathcal{C}$ is MDS;

$(2)$ any $k$ columns of $\boldsymbol{G}$ are $\mathbb{F}_{q}$-linearly independent;

$(3)$ any $n-k$ columns of  $\boldsymbol{H}$  are $\mathbb{F}_{q}$-linearly independent;

$(4)$ $\mathcal{C}^{\perp}$ is MDS.
\end{lemma}

%\begin{lemma}\label{NMDScondition}
%	{\rm(\cite{A25}, Lemma 3.7)}
 %An $[n, k]$ linear code $\mathcal{C}$ over $\mathbb{F}_{q}$ is NMDS if and only if a generator matrix $\boldsymbol{G}$ of $\mathcal{C}$ satisfies the following conditions:

%$(1)$ There exists $k$ linearly dependent columns in $\boldsymbol{G}$, i.e., $S(\mathcal{C}) \neq 0$ and $S\left(\mathcal{C}^{\perp}\right) \neq 0$.

%$(2)$ Any $k-1$ columns of $\boldsymbol{G}$ are $\mathbb{F}_{q}$-linearly
%independent, i.e., $S\left(\mathcal{C}^{\perp}\right) \leq 1$.

%$(3)$ Any $k+1$ columns of $\boldsymbol{G}$ are rank of $k$, i.e., $S(\mathcal{C}) \leq 1$.
%\end{lemma}
\begin{lemma}\label{Vandermonde}
{\rm(\cite{A9}, Lemma 6)} Let $x_1,\ldots,x_n$ be pairwise distinct elements of $\mathbb{F}_q$, then
	\begin{align*}
		\det\left(\begin{array}{ccccc}
			1 &  \ldots &  1 \\ 
			x_1  & \ldots &  x_n \\
			%a_1^2 &  \ldots & a_{n-1}^2 & a_n^2 \\
			\vdots &  \ddots &  \vdots \\
			x_1^{n-2} &  \ldots &  x_n^{n-2} \\
			x_1^n &  \ldots &  x_n^n 
		\end{array}\right)=\sum_{s=1}^{n}x_s\prod\limits_{1\leq i<j\leq n}(x_j-x_i).             
	\end{align*}
\end{lemma}

\begin{lemma}\label{paritycheckneedlemma} {\rm(\cite{A8}, Lemma 2.9)} Let $u_{i}=\prod\limits_{j=1, j \neq i}^{n}\left(\alpha_{i}-\alpha_{j}\right)^{-1}$ for $1 \leq i \leq n$.  Then for any subset $\left\{\alpha_{1}, \ldots, \alpha_{n}\right\}\subseteq\mathbb{F}_{q}$ with $n\geq 3$, we have
	$$\sum\limits_{i=1}^{n} u_{i} \alpha_{i}^{j}=\left\{\begin{array}{ll}
		0, &\text{if}\ 0 \leq j \leq n-2; \\
		1, & \text { if } j=n-1; \\
		\sum\limits_{i=1}^{n} \alpha_{i}, & \text { if } j=n. \\
	\end{array}\right.$$
\end{lemma}
 
If $\mathcal{C}$ is an $[n, k, n-k]$ NMDS code, then the weight distributions of $\mathcal{C}$ and its dual code are determined by the following lemma.
\begin{lemma}\label{NMDSweight}{\rm(\cite{A4}, Lemma 2)} Let $\mathcal{C}$ be an $[n, k, n-k]$ NMDS code over $\mathbb{F}_q$. Then $\mathcal{C}$ and $\mathcal{C}^{\perp}$ have the same number of
	minimum weight codewords, i.e., $A_{n-k}=A_{k}^{\perp}.$ Moreover, the
	weight distributions of $\mathcal{C}$ and $\mathcal{C}^{\perp}$ are given by
	\begin{equation*}\label{ANMDS}
		A_{n-k+s}=\binom{n}{k-s}\sum\limits_{j=0}^{s-1}(-1)^j\binom{n-k+s}{j}(q^{s-j}-1)\!+\!(-1)^{s}\binom{k}{s}A_{n-k}(1\leq s\leq k),
	\end{equation*}
	and
	 \begin{equation*}\label{DNMDS}
		A_{k+t}^{\perp}=\binom{n}{k+t}\sum\limits_{j=0}^{t-1}(-1)^j\binom{k+t}{j}(q^{t-j}-1)\!+\!(-1)^{t}\binom{n-k}{t}A_{k}^{\perp}(1\leq t\leq n-k),
	\end{equation*}
	respectively.
\end{lemma}

It's well-known that the subset sum problem is an $\mathbf{NP}$-complete problem. Given a subset $D$ of $\mathbb{F}_q$, a positive integer $k$ such that $1\leq k\leq |D|$ and  $b\in\mathbb{F}_q$, to determine the number $\# N(k,b,D)$ of $k$-element subsets $S\subseteq D$ such that 
\begin{align*}
\sum\limits_{a\in S}=b
\end{align*}
is just the subset sum problem. Especially, Li and Wan \cite{A35}  gave an explicit formula for $\# N(k,b,\mathbb{F}_q)$ and $\# N(k,b,\mathbb{F}_{q}^{*})$, respectively,  as the following
\begin{lemma}\label{subsetsum}
	{\rm(\cite{A36}, Theorem 1.2)} For any $b\in\mathbb{F}_{q}$, define $v(b)=\begin{cases}
		-1,&\text{if}\ b\neq 0;\\
		q-1,&\text{if}\ b=0,\\
	\end{cases}$ then		
	\begin{align*}
		\#N(k,b,\mathbb{F}_{q}^{*})=
		\frac{1}{q}\binom{q-1}{k}+(-1)^{k+\lfloor\frac{k}{p}\rfloor}\frac{v(b)}{q}\binom{\frac{q}{p}-1}{\lfloor\frac{k}{p}\rfloor},
	\end{align*}
	and 
	$$\# N(k,b,\mathbb{F}_{q})=\begin{cases}
		\frac{1}{q}\binom{q}{k},&\text{if}\ p\nmid k;\\
		\frac{1}{q}\binom{q}{k}+(-1)^{k+\frac{k}{p}}\frac{v(b)}{q}\binom{\frac{q}{p}}{\frac{k}{p}}, &\text{if}\ p\mid k.\\
	\end{cases}$$
\end{lemma}
\begin{lemma}\label{subsetsumneq0}
	{\rm(\cite{A19}, Remark 2.6)} If $k \geq 2$ and $D=\mathbb{F}_{q}^{*}$ or $\mathbb{F}_{q}$, then $\# N\left(k, b,D\right)=0$ if and only if both $2 | q$ and $(k, b) \in\{(2,0),(q-2,0)\}$.
\end{lemma}

By the proof of Lemma 6 in \cite{A10}, we have the following 
\begin{lemma}\label{subsetsum0}
	{\rm(\cite{A10})} $\# N(k-1, 0, \mathbb{F}_q^*) = 0$ if and only if $k \in \{3, q-2, q-1\}$ and $p=2$, or $k \in \{2, q-1\}$ and $p \neq 2$.
\end{lemma} 
\section{The parity-check matrix of $\text{EGRL}_{k,2,0}\left(\alpha,v,b\right)$}
In this section, for the EGRL code $\mathrm{EGRL}_{k,2,0}\left(\boldsymbol{\alpha},\boldsymbol{v},b\right)$, we give a parity-check matrix as the following  
\begin{theorem}\label{EGRLparitycheck}
Let $\mathbb{F}_q$ be the finite field of $q$ elements, where $q$ is a prime power. Let $\boldsymbol{\alpha}=\left(\alpha_{1}, \ldots, \alpha_{n}\right) \in \mathbb{F}_{q}^{n}$ with $\alpha_{i} \neq \alpha_{j}(i \neq j)$, $b\in\mathbb{F}_{q}^{*}$, $\boldsymbol{v}=$ $\left(v_{1}, \ldots, v_{n}\right) \in\left(\mathbb{F}_{q}^{*}\right)^{n}$ and $\boldsymbol{M}_{2\times 2}=(a_{ij})_{2\times 2}\in\mathrm{GL}_{2}\left(\mathbb{F}_{q}\right)$. Then $\mathrm{EGRL}_{k,2,0}\left(\boldsymbol{\alpha},\boldsymbol{v},b\right)(4\leq k\leq n-1)$ has the parity-check matrix 
	\begin{equation}\label{EGRLparitycheckmatrix}
		\boldsymbol{H}_{EGRL}=\begin{pmatrix}
			1& \cdots & 1&0& 0&-b^{-1}\sum\limits_{i=1}^{n}v_{i}\\
			\frac{u_{1}}{v_{1}} & \cdots & \frac{u_{n}}{v_{n}} &0& 0&0\\
			\frac{u_{1}}{v_{1}}\alpha_{1} & \cdots & \frac{u_{n}}{v_{n}}\alpha_{n} & 0& 0 & 0 \\
			\vdots & \ddots & \vdots & \vdots & \vdots& \vdots  \\
			\frac{u_{1}}{v_{1}}\alpha_{1}^{n-k-1} & \cdots & \frac{u_{n}}{v_{n}}\alpha_{n}^{n-k-1} & 0&0& 0 \\
			\frac{u_{1}}{v_{1}}\alpha_{1}^{n-k} & \cdots & \frac{u_{n}}{v_{n}}\alpha_{n}^{n-k} &  r_{11} &  r_{12}& 0\\
			\frac{u_{1}}{v_{1}}\alpha_{1}^{n-k+1} & \cdots & \frac{u_{n}}{v_{n}}\alpha_{n}^{n-k+1} & r_{21} &r_{22}& 0
		\end{pmatrix}_{(n-k+3)\times (n+3)},
	\end{equation}
	where 
	$$\boldsymbol{R}_{2\times 2}=\left(\begin{matrix}
		r_{11}&		r_{12}\\
		r_{21}&		r_{22}
	\end{matrix}\right)=\begin{pmatrix}
		0&-1\\
		-1&-\sum\limits_{i=1}^{n}\alpha_{i}
	\end{pmatrix}^{T}\left(\boldsymbol{M}_{2\times 2}^{T}\right)^{-1}.$$
\end{theorem}
{\bf Proof}. By Definition \ref{EGRLdefinition}, we know that $\mathrm{EGRL}_{k,2,0}\left(\boldsymbol{\alpha},\boldsymbol{v},b\right)$ has the generator matrix 
\begin{equation}\label{EGRL}
	\boldsymbol{G}_{EGRL}=\begin{pmatrix}    
		v_{1} & \cdots & v_{n} &0& 0&b
		\\
		v_{1}\alpha_{1} & \cdots & v_{n}\alpha_{n} & 0& 0 & 0 \\
		\vdots & \ddots & \vdots & \vdots & \vdots& \vdots\\
		v_{1}\alpha_{1}^{k-3} & \cdots & v_{n}\alpha_{n}^{k-3} & 0& 0 & 0\\
		v_{1}\alpha_{1}^{k-2} & \ldots & v_{n}\alpha_{n}^{k-2} &  a_{11} &  a_{12}& 0\\
		v_{1}\alpha_{1}^{k-1} & \cdots & v_{n}\alpha_{n}^{k-1} & a_{21} & a_{22}& 0 \\
	\end{pmatrix}_{k\times (n+3)}.
\end{equation} It's well-known that   $\boldsymbol{H}_{EGRL}$ is a parity-check matrix of $\mathrm{EGRL}_{k,2,0}\left(\boldsymbol{\alpha},\boldsymbol{v},b\right)$ if and only if $\mathrm{rank}(\boldsymbol{H}_{EGRL})=n+3-k$ and $\boldsymbol{G}_{EGRL}\boldsymbol{H}_{EGRL}^{T}=\boldsymbol{0}$. It's easy to prove that $\rank(\boldsymbol{H}_{EGRL})=n+3-k$, and so it's sufficient to check that $\boldsymbol{G}_{EGRL}\boldsymbol{H}_{EGRL}^{T}=\boldsymbol{0}$.

For the convenience, we set 
$$\boldsymbol{G}_{EGRL}=\begin{pmatrix}
	g_{0}\\
	g_{1}\\
	\vdots\\ 
	g_{k-3}\\
	g_{k-2}\\
	g_{k-1}
\end{pmatrix},\boldsymbol{H}_{EGRL}=\begin{pmatrix}
\tilde{h}\\
	h_{0}\\ 
	\vdots\\ 
	h_{n-k-1}\\
	h_{n-k}\\
	h_{n-k+1}
\end{pmatrix}.$$
Then it's easy to show that $$g_{i}\tilde{h}^{T}=\begin{cases}
\sum\limits_{s=1}^{n}v_{s}+b\cdot \left(-b^{-1}\sum\limits_{i=1}^{n}v_{i}\right),&\text{if}\ i=0;\\
\sum\limits_{s=1}^{n}u_{s}\alpha_{s}^{i},&\text{if}\ 1\leq i\leq k-1.
\end{cases}$$
It's obvious that $g_{0}\tilde{h}^{T}=0$. And by $k\leq n-1$, we have $k-1\leq n-2,$ and so by Lemma \ref{paritycheckneedlemma}, we know that $g_{i}\tilde{h}^{T}=0$ for any $1\leq i\leq k-1$. Furthermore, for any $0\leq i\leq k-1$, we have $g_{i}\tilde{h}^{T}=0$.

Next, we show that $g_{i}h_{j}^{T}=0$ for $0\leq i\leq k-1$ and $0\leq j\leq n-k+1$. In fact, it's easy to see that
$$g_{i}h_{j}^{T}=\begin{cases}
	\sum\limits_{s=1}^{n}u_{s}\alpha_{s}^{i+j},& \text{if}\ 0\leq i\leq k-3,0\leq j\leq n-k+1;\\
	&\ \ \  \text{or}\ k-2\leq i\leq k-1,0\leq j\leq n-k-1;\\
	\sum\limits_{s=1}^{n}u_{s}\alpha_{s}^{n-2}+a_{11}r_{11}+a_{12}r_{12},& \text{if}\ i=k-2,j=n-k;\\
	\sum\limits_{s=1}^{n}u_{s}\alpha_{s}^{n-1}+a_{11}r_{21}+a_{12}r_{22},& \text{if}\ i=k-2,j=n-k+1;\\
	\sum\limits_{s=1}^{n}u_{s}\alpha_{s}^{n-1}+a_{21}r_{11}+a_{22}r_{12},& \text{if}\ i=k-1,j=n-k;\\
	\sum\limits_{s=1}^{n}u_{s}\alpha_{s}^{n}+a_{21}r_{21}+a_{22}r_{22},& \text{if}\ i=k-1,j=n-k+1;\\
\end{cases}$$
and 
$$\footnotesize\left(\begin{matrix}
		a_{11}r_{11}+a_{12}r_{12}&a_{11}r_{21}+a_{12}r_{22}\\
		a_{21}r_{11}+a_{22}r_{12}&a_{21}r_{21}+a_{22}r_{22}
	\end{matrix}\right)=\left(\begin{matrix}
		a_{11}&		a_{12}\\
		a_{21}&		a_{22}
	\end{matrix}\right)\left(\begin{matrix}
		r_{11}&		r_{12}\\
		r_{21}&		r_{22}
	\end{matrix}\right)^{T} =\left(\boldsymbol{R}_{2\times 2}\boldsymbol{M}_{2\times 2}^{T}\right)^{T}=\begin{pmatrix}
		0&-1\\
		-1&-\sum\limits_{i=1}^{n}\alpha_{i}
	\end{pmatrix}.$$
And then by Lemma \ref{paritycheckneedlemma}, one can compute $g_{i}h_{j}^{T}=0$ for $0\leq i\leq k-1$ and $0\leq j\leq n-k+1$. 

From the above, we complete the proof of Theorem $\ref{EGRLparitycheck}$.
$\hfill\Box$

\section{The equivalent condition for $\text{EGRL}_{k,2,0}\left(\alpha,v,b\right)$ to be MDS }
In this section, we give an equivalent condition for $\mathrm{EGRL}_{k,2,0}\left(\boldsymbol{\alpha},\boldsymbol{v},b\right)$ to be MDS, namely, we prove the following

\begin{theorem}\label{ERLMDS}
Let $\mathbb{F}_q$ be the finite field of $q$ elements, where $q$ is a prime power. Let $\boldsymbol{\alpha}=\left(\alpha_{1}, \ldots, \alpha_{n}\right) \in \mathbb{F}_{q}^{n}$ with $\alpha_{i} \neq \alpha_{j}(i \neq j)$, $b\in\mathbb{F}_{q}^{*}$, $\boldsymbol{v}=$ $\left(v_{1}, \ldots, v_{n}\right) \in\left(\mathbb{F}_{q}^{*}\right)^{n}$ and $\boldsymbol{M}_{2\times 2}=(a_{ij})_{2\times 2}\in\mathrm{GL}_{2}\left(\mathbb{F}_{q}\right)$. Then $\mathrm{EGRL}_{k,2,0}\left(\boldsymbol{\alpha},\boldsymbol{v},b\right)$ is MDS if and only if the following two conditions hold simultaneously:
	
	$(1)$ $\boldsymbol{\alpha}=\left(\alpha_{1},\ldots,\alpha_{n}\right)\in\left(\mathbb{F}_{q}^{*}\right)^{n}$;
	
	$(2)$ for any subset $I_{m}\subseteq\left\{\alpha _{1},\ldots,\alpha _{n}\right\}$ with size $k-m(m=1,2)$, $a_{2j}\neq a_{1j}\sum\limits_{\alpha_{i}\in I_{m}} \alpha_{i} (j=1,2).$ 
	
	\iffalse$(3)$ for any subset $J\subseteq\left\{\alpha _{1},\ldots,\alpha _{n}\right\}$ with size $k-2$,
	$$a_{2j}\neq a_{1j}\sum\limits_{\alpha_{i}\in J} \alpha_{i},\ a_{2j}\neq a_{1j}\sum\limits_{\alpha_{i}\in I_{m}} \alpha_{i}.$$\fi
\end{theorem} 
{\bf Proof}. By Definition \ref{EGRLdefinition}, we know that $\mathrm{EGRL}_{k,2,0}\left(\boldsymbol{\alpha},\boldsymbol{v},b\right)$ and  $\mathrm{EGRL}_{k,2,0}\left(\boldsymbol{\alpha},\boldsymbol{1},b\right)$ have the generator matrix $\boldsymbol{G}_{EGRL}$ given by $(\ref{EGRL})$ and 
\begin{equation}\label{ERLgenerator}
	\boldsymbol{G}_{ERL}=\begin{pmatrix}
		1 & \cdots & 1 &0& 0&b \\
		\alpha_{1} & \cdots & \alpha_{n} & 0& 0 & 0 \\
		\vdots & \ddots & \vdots & \vdots & \vdots& \vdots  \\
		\alpha_{1}^{k-3} & \cdots & \alpha_{n}^{k-3} & 0& 0 & 0 \\
		\alpha_{1}^{k-2} & \ldots & \alpha_{n}^{k-2} &  a_{11} &  a_{12}& 0 \\
		\alpha_{1}^{k-1} & \cdots & \alpha_{n}^{k-1} & a_{21} & a_{22}& 0
	\end{pmatrix}_{k\times (n+3)},
\end{equation}
respectively. Note that $$\boldsymbol{G}_{EGRL}=\boldsymbol{G}_{ERL}\cdot\mathrm{diag}\left\{v_{1},\ldots,v_{n},1,1,1\right\}\iffalse\begin{pmatrix}
	v_{1}& & & & & \\
	&\ddots& & & & \\
	& &v_{n}& & & \\
	& & &1& & \\
	& & & &1& \\
	& & & & &1\\
\end{pmatrix}\fi,$$ 
And then by Lemma \ref{monomiallyequivalent},   $\mathrm{EGRL}_{k,2,0}\left(\boldsymbol{\alpha},\boldsymbol{v},b\right)$ and $\mathrm{EGRL}_{k,2,0}\left(\boldsymbol{\alpha},\boldsymbol{1},b\right)$ are monomially equivalent. Thus, we only focus on  $\mathrm{EGRL}_{k,2,0}\left(\boldsymbol{\alpha},\boldsymbol{1},b\right)$.
 
Note that any $k\times k$ submatrices of $\boldsymbol{G}_{ERL}$ has one of the following forms,
$$
\boldsymbol{D}_{1}=\begin{pmatrix}
	1 & \cdots & 1\\
	\alpha_{i_1} & \cdots & \alpha_{i_{k}}\\
	\vdots & \ddots & \vdots \\
	
	\alpha_{i_1}^{k-3} & \cdots & \alpha_{i_{k}}^{k-3}\\ 
	\alpha_{i_1}^{k-2} & \ldots & \alpha_{i_{k}}^{k-2}\\
	\alpha_{i_1}^{k-1} & \cdots & \alpha_{i_{k}}^{k-1}
\end{pmatrix}_{k\times k},\ 
\boldsymbol{D}_{2}=\begin{pmatrix}
	1 & \cdots & 1&b\\
	\alpha_{i_1} & \cdots & \alpha_{i_{k-1}}&0\\
	\vdots & \ddots & \vdots& \vdots \\
	
	\alpha_{i_1}^{k-3} & \cdots & \alpha_{i_{k-1}}^{k-3}&0\\ 
	\alpha_{i_1}^{k-2} & \ldots & \alpha_{i_{k-1}}^{k-2}&0\\
	\alpha_{i_1}^{k-1} & \cdots & \alpha_{i_{k-1}}^{k-1}&0
\end{pmatrix}_{k\times k},$$
$$\boldsymbol{D}_{3}=\begin{pmatrix}
1 & \cdots & 1&0&0&b\\
\alpha_{i_1} & \cdots & \alpha_{i_{k-3}}&0&0&0\\
\vdots & \ddots & \vdots& \vdots&\vdots& \vdots \\

\alpha_{i_1}^{k-3} & \cdots & \alpha_{i_{k-3}}^{k-3}&0&0&0\\ 
\alpha_{i_1}^{k-2} & \ldots & \alpha_{i_{k-3}}^{k-2}&a_{11}&a_{12}&0\\
\alpha_{i_1}^{k-1} & \cdots & \alpha_{i_{k-3}}^{k-1}&a_{21}&a_{22}&0
\end{pmatrix}_{k\times k},
$$
$$\boldsymbol{D}_{4,s}=\begin{pmatrix}
	1 & \cdots & 1&0\\
	\alpha_{i_1} & \cdots & \alpha_{i_{k-1}}&0\\
	\vdots & \ddots & \vdots& \vdots \\
	
	\alpha_{i_1}^{k-3} & \cdots & \alpha_{i_{k-1}}^{k-3}&0\\ 
	\alpha_{i_1}^{k-2} & \ldots & \alpha_{i_{k-1}}^{k-2}&a_{1s}\\
	\alpha_{v1}^{k-1} & \cdots & \alpha_{i_{k-1}}^{k-1}&a_{2s}
\end{pmatrix}_{k\times k},\ \boldsymbol{D}_{5,s}=\begin{pmatrix}
1 & \cdots & 1&0&b\\
\alpha_{i_1} & \cdots & \alpha_{i_{k-2}}&0&0\\
\vdots & \ddots & \vdots& \vdots& \vdots \\
\alpha_{i_1}^{k-3} & \cdots & \alpha_{i_{k-2}}^{k-3}&0&0\\ 
\alpha_{i_1}^{k-2} & \ldots & \alpha_{i_{k-2}}^{k-2}&a_{1s}&0\\
\alpha_{i_1}^{k-1} & \cdots & \alpha_{i_{k-2}}^{k-1}&a_{2s}&0
\end{pmatrix}_{k\times k}
(s=1,2).
$$  
By Lemma \ref{MDSdefinition}, we know that $\mathrm{EGRL}_{k,2,0}\left(\boldsymbol{\alpha},\boldsymbol{1},b\right)$ is MDS if and only if any $k$ columns of $\boldsymbol{G}_{ERL}$ are $\mathbb{F}_{q}$-linearly independent, i.e., any $k\times k$ submatrices of $\boldsymbol{G}_{ERL}$ is nonsingular over $\mathbb{F}_{q}$, which means that any of the following five determinants is not equal to zero,
$$
\det (\boldsymbol{D}_1)=\prod\limits_{1\leq j<l\leq k}(\alpha_{i_l}-\alpha_{i_j}),
$$
$$
\det (\boldsymbol{D}_{2})=(-1)^{k+1}b\cdot\prod\limits_{t=1}^{k-1}\alpha_{i_t}\cdot\prod\limits_{1\leq j<l\leq k-1}(\alpha_{i_l}-\alpha_{i_j}),
$$
$$\det(\boldsymbol{D}_{3})=(-1)^{k+1}b\cdot\det(\boldsymbol{M}_{2\times 2})\cdot\prod\limits_{t=1}^{k-3}\alpha_{i_t}\cdot\prod\limits_{1\leq j<l\leq k-3}(\alpha_{i_l}-\alpha_{i_j}),$$
$$
\det (\boldsymbol{D}_{4,s})=(a_{2s}-a_{1s}\sum\limits_{j=1}^{k-1} \alpha_{i_j})\prod\limits_{1\leq j<l\leq k-1}(\alpha_{i_l}-\alpha_{i_j}),
$$ 
$$
\det (\boldsymbol{D}_{5,s})=(-1)^{k+1}b\cdot\prod\limits_{t=1}^{k-2}\alpha_{i_t}\cdot(a_{2s}-a_{1s}\sum\limits_{j=1}^{k-2} \alpha_{i_j})\cdot\prod\limits_{1\leq j<l\leq k-2}(\alpha_{i_l}-\alpha_{i_j}).
$$ 
Note that $b\in\mathbb{F}_{q}^{*}$, $\boldsymbol{M}_{2\times 2}\in\mathrm{GL}_{2}\left(\mathbb{F}_{q}\right)$ and  $\boldsymbol{\alpha}=\left(\alpha_{1}, \ldots, \alpha_{n}\right) \in \mathbb{F}_{q}^{n}$ with $\alpha_{i} \neq \alpha_{j}(i \neq j)$, thus we know that $\det(\boldsymbol{M}_{2\times 2})$, $\prod\limits_{1\leq j<l\leq k}(\alpha_{i_l}-\alpha_{i_j})$, $\prod\limits_{1\leq j<l\leq k-1}(\alpha_{i_l}-\alpha_{i_j})$,  $\prod\limits_{1\leq j<l\leq k-2}(\alpha_{i_l}-\alpha_{i_j})$ and $\prod\limits_{1\leq j<l\leq k-3}(\alpha_{i_l}-\alpha_{i_j})$ are not equal to zero, furthermore, $\mathrm{EGRL}_{k,2,0}\left(\boldsymbol{\alpha},\boldsymbol{1},b\right)$ is MDS if and only if the following conditions hold simultaneously: 

$(1)$ for any subset $I_{m}\subseteq\left\{\alpha _{1},\ldots,\alpha _{n}\right\}$ with size $k-m(m=1,2,3)$,  $\prod\limits_{\alpha_{i}\in I_{m}}\alpha_{i_t}\neq 0,$ i.e., $$\boldsymbol{\alpha}=\left(\alpha_{1},\ldots,\alpha_{n}\right)\in\left(\mathbb{F}_{q}^{*}\right)^{n};$$

$(2)$ for any subset $I_{m}\subseteq\left\{\alpha _{1},\ldots,\alpha _{n}\right\}$ with size $k-m(m=1,2)$, 
$$a_{2j}- a_{1j}\sum\limits_{\alpha_{i}\in I_{m}} \alpha_{i}\neq 0 (j=1,2),$$
i.e., $a_{2j}\neq a_{1j}\sum\limits_{\alpha_{i}\in I_{m}} \alpha_{i} (j=1,2).$ 
\iffalse$(2)$ for any subset $I\subseteq\left\{\alpha _{1},\ldots,\alpha _{n}\right\}$ with size $k-1$,  $$a_{11}-a_{21}\sum\limits_{\alpha_{i}\in I} \alpha_{i}\neq 0,\ a_{22}-a_{12}\sum\limits_{\alpha_{i}\in I} \alpha_{i}\neq 0,$$
i.e., $$a_{21}\neq a_{11}\sum\limits_{\alpha_{i}\in I} \alpha_{i},\ a_{22}\neq a_{12}\sum\limits_{\alpha_{i}\in I} \alpha_{i}.$$

$(2)$ for any subset $J\subseteq\left\{\alpha _{1},\ldots,\alpha _{n}\right\}$ with size $k-2$,  $$a_{21}-a_{11}\sum\limits_{\alpha_{i}\in J} \alpha_{i}\neq 0,\ a_{22}-a_{12}\sum\limits_{\alpha_{i}\in J} \alpha_{i}\neq 0,$$
i.e., $$a_{21}\neq a_{11}\sum\limits_{\alpha_{i}\in J} \alpha_{i},\ a_{22}\neq a_{12}\sum\limits_{\alpha_{i}\in J} \alpha_{i}.$$\fi

This completes the proof of Theorem $\ref{ERLMDS}$.

 $\hfill\Box$
 
\begin{example}\label{ERLMDSexample1}
	Let $(q,n,k)=(13,5,5),\boldsymbol{v}=\boldsymbol{1}\in\mathbb{F}_{q}^{n}$ and $ \boldsymbol{\alpha}=\left(1,2,7,8,9\right)$. Then we immediately know  that the corresponding code $\mathrm{EGRL}_{k,2,0}\left(\boldsymbol{\alpha},\boldsymbol{1},b\right)$ has the following generator matrix
	$$\boldsymbol{G}_{ERL}=\left(\begin{matrix}
		1&		1&1&		1&		1&		0&		0&		b\\
		1&2&7&8&9&		0&0&0\\
		1&4&10&12&3&0&0&0\\
		1&8&5&5&1&1&		1&		0\\
		1&3&9&1&9&1&		2&		0\\
	\end{matrix} \right)_{5\times 8},$$
where $b\in\mathbb{F}_{q}^{*}$. By directly calculating, we obtain the following two tables.
	\begin{table}[H]
	\centering
	\footnotesize 
	\label{table_example1.1}
	\begin{tabular}{|c|c|c|c|}
		\hline	
		$I_{1}$&$\sum\limits_{\alpha_{i}\in I_{1}} \alpha_{i_j}$&$ a_{11}\sum\limits_{\alpha_{i}\in I_{1}} \alpha_{i}\neq a_{21}$ &$a_{12}\sum\limits_{\alpha_{i}\in I_{1}} \alpha_{i}\neq a_{22}$\\
		\hline
		$\left\{1,2,7,8\right\}$&$5$&$5\neq 1$ &$5\neq 2$ \\
		\hline
		$\left\{1,2,7,9\right\}$&$6$&$6\neq 1$ &$6\neq 2$\\
		\hline
		$\left\{1,2,8,9\right\}$&$7$&$7\neq 1$ &$7\neq 2$ \\
		\hline
		$\left\{1,7,8,9\right\}$&$12$& $12\neq 1$&$12\neq 2$ \\
		\hline
		$\left\{2,7,8,9\right\}$&$0$& $0\neq 1$&$0\neq 2$\\		\hline
	\end{tabular}
\end{table}
%\begin{table}[H]
%	\centering
%	\footnotesize 
%	\label{table_example1}
%	\begin{tabular}{|c|c|c|c|c|c|}
%		\hline	
%		$I$&$\left\{1,2,7,8\right\}$&$\left\{1,2,7,9\right\}$&$\left\{1,2,8,9\right\}$&$\left\{1,7,8,9\right\}$ &$\left\{2,7,8,9\right\}$\\
%		\hline
%		$\sum\limits_{j=1}^{k-1} \alpha_{i_j}$&$5$&$6$ &$7$&$12$&$0$ \\
%		
%		\hline
%		$ a_{11}\sum\limits_{\alpha_{i}\in I} \alpha_{i}\neq a_{21}$&$5\neq 1$& $6\neq 1$&$7\neq 1$&$12\neq 1$&$0\neq 1$ \\
%		\hline
%		$a_{12}\sum\limits_{\alpha_{i}\in I} \alpha_{i}\neq a_{22}$&$5\neq 2$& $6\neq 2$&$7\neq 2$&$12\neq 2$&$0\neq 2$\\
%		\hline
%	\end{tabular}
%\end{table}	
%\begin{table}[H]
%	\centering
%	\footnotesize 
%	\label{table_example2}
%	\begin{tabular}{|c|c|c|c|c|c|c|c|c|c||c|c|}
%		\hline	
%		$I$&$\left\{1,2,7\right\}$&$\left\{1,2,7\right\}$&$\left\{1,2,7\right\}$&$\left\{1,2,7\right\}$&$\left\{1,2,7\right\}$&$\left\{1,2,7\right\}$&$\left\{1,2,7\right\}$&$\left\{1,2,7\right\}$&$\left\{1,2,7\right\}$&$\left\{1,2,7\right\}$&\\
%		\hline
%		$\sum\limits_{j=1}^{k-1} \alpha_{i_j}$&$5$&$6$ &$7$&$12$&$0$ \\
%		
%		\hline
%		$ a_{11}\sum\limits_{\alpha_{i}\in I} \alpha_{i}\neq a_{21}$&$5\neq 1$& $6\neq 1$&$7\neq 1$&$12\neq 1$&$0\neq 1$ \\
%		\hline
%		$a_{12}\sum\limits_{\alpha_{i}\in I} \alpha_{i}\neq a_{22}$&$5\neq 2$& $6\neq 2$&$7\neq 2$&$12\neq 2$&$0\neq 2$\\
%		\hline
%	\end{tabular}
%\end{table}	
	\begin{table}[H]
	\centering
	\footnotesize 
	\label{table_example1.2}
	\begin{tabular}{|c|c|c|c|}
		\hline	
		$I_{2}$&$\sum\limits_{\alpha_{i}\in I_{2}} \alpha_{i}$&$ a_{11}\sum\limits_{\alpha_{i}\in I_{2}} \alpha_{i}\neq a_{21}$ &$a_{12}\sum\limits_{\alpha_{i}\in I_{2}} \alpha_{i}\neq a_{22}$\\
		\hline
		$\left\{1,2,7\right\}$&$10$&$10\neq 1$ &$10\neq 2$ \\
		\hline
		$\left\{1,2,8\right\}$&$11$&$11\neq 1$ &$11\neq 2$\\
		\hline
		$\left\{1,2,9\right\}$&$12$&$12\neq 1$ &$12\neq 2$ \\
		\hline
		$\left\{1,7,8\right\}$&$3$& $3\neq 1$&$3\neq 2$ \\
		\hline
		$\left\{1,7,9\right\}$&$4$& $4\neq 1$&$4\neq 2$\\
		\hline
		$\left\{1,8,9\right\}$&$5$& $5\neq 1$&$5\neq 2$\\
		\hline
		$\left\{2,7,8\right\}$&$4$& $4\neq 1$&$4\neq 2$\\
		\hline
		$\left\{2,7,9\right\}$&$5$& $5\neq 1$&$5\neq 2$\\
		\hline
		$\left\{2,8,9\right\}$&$6$& $6\neq 1$&$6\neq 2$\\
		\hline
		$\left\{7,8,9\right\}$&$11$& $11\neq 1$&$11\neq 2$\\
		\hline
	\end{tabular}
	\end{table}
Now by Theorem \ref{ERLMDS} and the above two tables,  $\mathrm{EGRL}_{k,2,0}\left(\boldsymbol{\alpha},\boldsymbol{1},b\right)$ is MDS. Furthermore, based on the Magma programe, $\mathrm{EGRL}_{k,2,0}\left(\boldsymbol{\alpha},\boldsymbol{1},b\right)$ is a $\mathbb{F}_{13}$-linear code with the parameters $\left[8,5,4\right].$
\end{example}

%By Theorem $\ref{ERLMDSNMDS}$, we can immediately the following corollary
%\begin{corollary}
%	Let $\boldsymbol{\alpha}=\left\{\alpha_{1},\ldots,\alpha_{n}\right\}\in\mathbb{F}_{q}^{n}$ with $\alpha_i\neq \alpha_j(i\neq j)$, and $\boldsymbol{M}_{2\times 2}=(a_{ij})\in\mathrm{GL}_{2}(\mathbb{F}_{q})$, then $\mathrm{EGRL}_{k,2,0}\left(\boldsymbol{\alpha},\boldsymbol{1},b\right)$ is NMDS if and only if the following conditions hold simultaneously:

%$(1)$ $0\in \left\{\alpha _{1},\ldots,\alpha _{n}\right\}\in \mathbb{F}_{q}^{n}.$

%$(2)$ there exists some subset $I\subseteq\left\{\alpha _{1},\ldots,\alpha _{n}\right\}$ with size $k-1$,
%$$a_{11}= a_{21}\sum\limits_{j=1}^{k-1} \alpha_{i_j},\ a_{12}= a_{22}\sum\limits_{j=1}^{k-1} \alpha_{i_j}.$$

%$(2)$ there exists some subset $J\subseteq\left\{\alpha _{1},\ldots,\alpha _{n}\right\}$ with size $k-2$,
%$$a_{11}= a_{21}\sum\limits_{j=1}^{k-2} \alpha_{i_j},\ a_{12}= a_{22}\sum\limits_{j=1}^{k-2} \alpha_{i_j}.$$
%\end{corollary}
\section{The equivalent condition for $\text{EGRL}_{k,2,0}^{\perp}\left(\alpha,v,b\right)$ to be AMDS}
In this section, we give an equivalent condition for $\mathrm{EGRL}_{k,2,0}^{\perp}\left(\boldsymbol{\alpha},\boldsymbol{v},b\right)$ to be AMDS, namely, we prove the following

\begin{theorem}
\label{ERLAMDS}
Let $\mathbb{F}_q$ be the finite field of $q$ elements, where $q$ is a prime power. Let $\boldsymbol{\alpha}=\left(\alpha_{1}, \ldots, \alpha_{n}\right) \in \mathbb{F}_{q}^{n}$ with $\alpha_{i} \neq \alpha_{j}(i \neq j)$, $b\in\mathbb{F}_{q}^{*}$, $\boldsymbol{v}=$ $\left(v_{1}, \ldots, v_{n}\right) \in\left(\mathbb{F}_{q}^{*}\right)^{n}$ and $\boldsymbol{M}_{2\times 2}=(a_{ij})_{2\times 2}\in\mathrm{GL}_{2}\left(\mathbb{F}_{q}\right)$. Then $\mathrm{EGRL}_{k,2,0}^{\perp}\left(\boldsymbol{\alpha},\boldsymbol{v},b\right)$ is AMDS if and only if the following two conditions hold simultaneously:

$(1)$ $\boldsymbol{\alpha}=\left(\alpha_{1},\ldots,\alpha_{n}\right)\in\left(\mathbb{F}_{q}^{*}\right)^{n}$;

$(2)$ there exists some subset $I_{m}\subseteq\left\{\alpha _{1},\ldots,\alpha _{n}\right\}$ with size $k-m(m=1,2)$, such that 
$$a_{21}= a_{11}\sum\limits_{\alpha_{i}\in I_{m}} \alpha_{i},\ \text{or}\ a_{22}= a_{12}\sum\limits_{\alpha_{i}\in I_{m}} \alpha_{i}.$$

\iffalse$(3)$ there exists some subset $J\subseteq\left\{\alpha _{1},\ldots,\alpha _{n}\right\}$ with size $k-2$, such that 
$$a_{21}= a_{11}\sum\limits_{\alpha_{i}\in J} \alpha_{i},\ \text{or}\   a_{22}= a_{12}\sum\limits_{\alpha_{i}\in J} \alpha_{i}.$$\fi
\end{theorem}
{\bf Proof}. By  Lemma \ref{monomiallyequivalent} and the proof of Theorem $\ref{ERLMDS}$, we know that $\mathrm{EGRL}_{k,2,0}\left(\boldsymbol{\alpha},\boldsymbol{v},b\right)$ and $\mathrm{EGRL}_{k,2,0}\left(\boldsymbol{\alpha},\boldsymbol{1},b\right)$ are monomially equivalent. Then we only focus on $\mathrm{EGRL}_{k,2,0}^{\perp}\left(\boldsymbol{\alpha},\boldsymbol{1},b\right)$. It's easy to know that $\boldsymbol{G}_{ERL}$ given by $(\ref{ERLgenerator})$ is a parity-check matrix of $\mathrm{EGRL}_{k,2,0}^{\perp}\left(\boldsymbol{\alpha},\boldsymbol{1},b\right)$, and then by the definition,   $\mathrm{EGRL}_{k,2,0}^{\perp}\left(\boldsymbol{\alpha},\boldsymbol{1},b\right)$ is AMDS if  and only if it has parameters $\left[n+3,n+3-k,k\right]$, i.e., $$d\left(\mathrm{EGRL}_{k,2,0}^{\perp}\left(\boldsymbol{\alpha},\boldsymbol{1},b\right)\right)=k.$$
Now by the definition, the minimum distance of  $\mathrm{EGRL}_{k,2,0}^{\perp}\left(\boldsymbol{\alpha},\boldsymbol{1},b\right)$ equals to $k$ if and only if the following two statements hold simultaneously:

(i) any $k-1$ columns of  $\boldsymbol{G}_{ERL}$ is $\mathbb{F}_{q}$-linearly independent;

(ii) there exists $k$ columns of $\boldsymbol{G}_{ERL}$ which are $\mathbb{F}_{q}$-linearly dependent.\\
And then we only need to prove that the above two statements (i) and (ii) hold simultaneously if and only if  the following two conditions hold simultaneously:

$(1)$ $\boldsymbol{\alpha}=\left(\alpha_{1},\ldots,\alpha_{n}\right)\in\left(\mathbb{F}_{q}^{*}\right)^{n}$;

$(2)$ there exists some subset $I_{m}\subseteq\left\{\alpha _{1},\ldots,\alpha _{n}\right\}$ with size $k-m(m=1,2)$, such that 
$$a_{21}= a_{11}\sum\limits_{\alpha_{i}\in I_{m}} \alpha_{i},\ \text{or}\  a_{22}= a_{12}\sum\limits_{\alpha_{i}\in I_{m}} \alpha_{i}.$$

\textbf{For (i)}, note that (i) holds if and only if any $k\times (k-1)$ submatrices of $\boldsymbol{G}_{ERL}$ exists a $(k-1)\times (k-1)$ non-zero minor. It's easy to know that any $k\times (k-1)$ submatrices of $\boldsymbol{G}_{ERL}$ has one of the following forms,
$$\boldsymbol{F}_{1}=\begin{pmatrix}
	1 & \cdots & 1\\
	\alpha_{i_1} & \cdots & \alpha_{i_{k-1}}\\
	\vdots & \ddots & \vdots \\
	
	\alpha_{i_1}^{k-3} & \cdots & \alpha_{i_{k-1}}^{k-3}\\ 
	\alpha_{i_1}^{k-2} & \ldots & \alpha_{i_{k-1}}^{k-2}\\
	\alpha_{i_1}^{k-1} & \cdots & \alpha_{i_{k-1}}^{k-1}
\end{pmatrix}_{k\times (k-1)},\boldsymbol{F}_{2}=\begin{pmatrix}
1 & \cdots & 1&b\\
\alpha_{i_1} & \cdots & \alpha_{i_{k-2}}&0\\
\vdots & \ddots & \vdots& \vdots \\

\alpha_{i_1}^{k-3} & \cdots & \alpha_{i_{k-2}}^{k-3}&0\\ 
\alpha_{i_1}^{k-2} & \ldots & \alpha_{i_{k-2}}^{k-2}&0\\
\alpha_{i_1}^{k-1} & \cdots & \alpha_{i_{k-2}}^{k-1}&0
\end{pmatrix}_{k\times (k-1)},$$
$$  \boldsymbol{F}_{3,s}=\begin{pmatrix}
	1 & \cdots & 1&0\\
	\alpha_{i_1} & \cdots & \alpha_{i_{k-2}}&0\\
	\vdots & \ddots & \vdots& \vdots \\
	
	\alpha_{i_1}^{k-3} & \cdots & \alpha_{i_{k-2}}^{k-3}&0\\ 
	\alpha_{i_1}^{k-2} & \ldots & \alpha_{i_{k-2}}^{k-2}&a_{1s}\\
	\alpha_{i_1}^{k-1} & \cdots & \alpha_{i_{k-2}}^{k-1}&a_{2s}
\end{pmatrix}_{k\times (k-1)},\boldsymbol{F}_{4,s}=\begin{pmatrix}
1 & \cdots & 1&0&b\\
\alpha_{i_1} & \cdots & \alpha_{i_{k-3}}&0&0\\
\vdots & \ddots & \vdots& \vdots \\

\alpha_{i_1}^{k-3} & \cdots & \alpha_{i_{k-3}}^{k-3}&0&0\\ 
\alpha_{i_1}^{k-2} & \ldots & \alpha_{i_{k-3}}^{k-2}&a_{1s}&0\\
\alpha_{i_1}^{k-1} & \cdots & \alpha_{i_{k-3}}^{k-1}&a_{2s}&0
\end{pmatrix}_{k\times (k-1)}(s=1,2),$$
$$\boldsymbol{F}_{5}=\begin{pmatrix}
1 & \cdots & 1&0&0&b\\
\alpha_{i_1} & \cdots & \alpha_{i_{k-4}}&0&0&0\\
\vdots & \ddots & \vdots& \vdots \\

\alpha_{i_1}^{k-3} & \cdots & \alpha_{i_{k-4}}^{k-3}&0&0&0\\ 
\alpha_{i_1}^{k-2} & \ldots & \alpha_{i_{k-4}}^{k-2}&a_{11}&a_{12}&0\\
\alpha_{i_1}^{k-1} & \cdots & \alpha_{i_{k-4}}^{k-1}&a_{21}&a_{22}&0
\end{pmatrix}_{k\times (k-1)}.$$ 
 
For $\boldsymbol{F}_{1}$, note that the matrix
$$\boldsymbol{K}_{1}=\begin{pmatrix}
	1 & \cdots & 1\\
	\alpha_{i_1} & \cdots & \alpha_{i_{k-1}}\\
	\vdots & \ddots & \vdots \\
	
	\alpha_{i_1}^{k-3} & \cdots & \alpha_{i_{k-1}}^{k-3}\\ 
	\alpha_{i_1}^{k-2} & \ldots & \alpha_{i_{k-1}}^{k-2}
\end{pmatrix}_{(k-1)\times (k-1)}$$ given by deleting the last row of $\boldsymbol{F}_{1}$ is the Vandermonde martix, then $\det(\boldsymbol{K}_{1})$ is a $(k-1)\times (k-1)$ non-zero minors of $\boldsymbol{F}_{1}$. 

For $\boldsymbol{F}_{2}$, we consider the matrix $$\boldsymbol{K}_{2}=\begin{pmatrix}
	1 & \cdots & 1&b\\
	\alpha_{i_1} & \cdots & \alpha_{i_{k-2}}&0\\
	\vdots & \ddots & \vdots& \vdots \\
	
	\alpha_{i_1}^{k-3} & \cdots & \alpha_{i_{k-2}}^{k-3}&0\\ 
	\alpha_{i_1}^{k-2} & \ldots & \alpha_{i_{k-2}}^{k-2}&0
\end{pmatrix}_{(k-1)\times (k-1)}$$ given by deleting the last row of  $\boldsymbol{F}_{2}$. Note that
$$\det(\boldsymbol{K_{2}})=b\prod\limits_{t=1}^{k-2}\alpha_{i_{t}}\cdot\prod\limits_{1 \leq j< l\leq k-2}(\alpha_{i_{l}}-\alpha_{i_{j}}),$$  
hence, $\mathrm{det}(\boldsymbol{K}_2)$ is a $(k-1)\times (k-1)$ non-zero minor of $\boldsymbol{F}_{2}$ if and only if for any $(k-2)$-elements subset $I_{2}\subseteq\left\{\alpha _{1},\ldots,\alpha _{n}\right\}$, $\prod\limits_{\alpha_{i_{t}}\in I_{2}} \alpha_{i_{t}}\neq 0$. 

For $\boldsymbol{F}_{3,s}$, note that $\boldsymbol{M}_{2\times 2}\in\mathrm{GL}_{2}\left(\mathbb{F}_{q}\right)$, it means that $a_{1s}$ and $a_{2s}$ can not be zero, simultaneously. Without loss of generality, we set $a_{1s}\neq 0$ and consider the matrix $$\boldsymbol{K}_{3,s}=\begin{pmatrix}
	1 & \cdots & 1&0\\
	\alpha_{i_1} & \cdots & \alpha_{i_{k-2}}&0\\
	\vdots & \ddots & \vdots& \vdots \\
	
	\alpha_{i_1}^{k-3} & \cdots & \alpha_{i_{k-2}}^{k-3}&0\\ 
	\alpha_{i_1}^{k-2} & \ldots & \alpha_{i_{k-2}}^{k-2}&a_{1s}
\end{pmatrix}_{(k-1)\times (k-1)}$$ given by deleting the last row of  $\boldsymbol{F}_{3,s}$. Note that 
$$\det(\boldsymbol{K_{3,s}})=a_{1s}\prod\limits_{1 \leq j< l\leq k-2}(\alpha_{i_{l}}-\alpha_{i_{j}})\neq 0,$$  
hence, $\det(\boldsymbol{K}_{3,s})$ is a $(k-1)\times (k-1)$ non-zero minor of $\boldsymbol{F}_{3,s}$. 

For $\boldsymbol{F}_{4,s}$, note that  $\boldsymbol{M}_{2\times 2}\in\mathrm{GL}_{2}\left(\mathbb{F}_{q}\right)$, it means that $a_{1s}$ and $a_{2s}$ can not be zero, simultaneously. Without loss of generality, we set $a_{1s}\neq 0$ and consider the matrix $$\boldsymbol{K}_{4,s}=\begin{pmatrix}
1 & \cdots & 1&0&b\\
\alpha_{i_1} & \cdots & \alpha_{i_{k-3}}&0&0\\
\vdots & \ddots & \vdots& \vdots \\

\alpha_{i_1}^{k-3} & \cdots & \alpha_{i_{k-3}}^{k-3}&0&0\\ 
\alpha_{i_1}^{k-2} & \ldots & \alpha_{i_{k-3}}^{k-2}&a_{1s}&0
\end{pmatrix}_{(k-1)\times (k-1)}$$ given by deleting the last row of  $\boldsymbol{F}_{4,s}$. Note that
$$\det(\boldsymbol{K_{4,s}})=(-1)^{k}ba_{1s}\cdot \prod\limits_{t=1}^{k-3}\alpha_{i_{t}}\cdot\prod\limits_{1 \leq j< l\leq k-3}(\alpha_{i_{l}}-\alpha_{i_{j}}),$$  
hence, $\mathrm{det}(\boldsymbol{K}_{4,s})$ is a $(k-1)\times (k-1)$ non-zero minor of $\boldsymbol{F}_{4,s}$ if and only if for any $(k-3)$-elements subset $I_{3}\subseteq\left\{\alpha _{1},\ldots,\alpha _{n}\right\}$, $\prod\limits_{\alpha_{i_{t}}\in I_{3}} \alpha_{i_{t}}\neq 0$. 

For $\boldsymbol{F}_{5}$, we consider the matrix $$\boldsymbol{K}_{5}=\begin{pmatrix}

1 & \cdots & 1&0&0&b\\
\alpha_{i_1} & \cdots & \alpha_{i_{k-4}}&0&0&0\\
\vdots & \ddots & \vdots& \vdots \\

\alpha_{i_1}^{k-4} & \cdots & \alpha_{i_{k-4}}^{k-4}&0&0&0\\ 
\alpha_{i_1}^{k-2} & \ldots & \alpha_{i_{k-4}}^{k-2}&a_{11}&a_{12}&0\\
\alpha_{i_1}^{k-1} & \cdots & \alpha_{i_{k-4}}^{k-1}&a_{21}&a_{22}&0
\end{pmatrix}_{(k-1)\times (k-1)}$$ given by deleting the $(k-2)$-th row of  $\boldsymbol{F}_{5}$. Note that
$$\det(\boldsymbol{K_{5}})=b\cdot\det(\boldsymbol{M}_{2\times 2})\cdot\prod\limits_{t=1}^{k-4}\alpha_{i_{t}}\cdot\prod\limits_{1 \leq j< l\leq k-4}(\alpha_{i_{l}}-\alpha_{i_{j}}),$$  
hence, $\mathrm{det}(\boldsymbol{K}_5)$ is a $(k-1)\times (k-1)$ non-zero minor of $\boldsymbol{F}_{5}$ if and only if for any $(k-4)$-elements subset $I_{4}\subseteq\left\{\alpha _{1},\ldots,\alpha _{n}\right\}$, $\prod\limits_{\alpha_{i_{t}}\in I_{4}} \alpha_{i_{t}}\neq 0$.

From the above discussions, we know that the statement (i) holds if and only if for any $(k-s)$-elements subset $I_{s}\subseteq\left\{\alpha _{1},\ldots,\alpha _{n}\right\} (s=2,3,4)$, $\prod\limits_{\alpha_{i_{t}}\in I_{s}} \alpha_{i_{t}}\neq 0$, i.e., $\alpha_{i}\in \mathbb{F}_{q}^{*}(i=1,\ldots,n)$, namely, Theorem $\ref{ERLAMDS}$ $(1)$ holds.

\textbf{For (ii)}, note that $\mathrm{EGRL}_{k,2,0}\left(\boldsymbol{\alpha},\boldsymbol{1},b\right)$ is MDS if and only if any $k$ columns of $\boldsymbol{G}_{ERL}$ are $\mathbb{F}_{q}$-linearly
independent. And so  by Theorem $\ref{ERLMDS}$, the statement (ii) holds if and only if there exists some subset $I_{m}\subseteq\left\{\alpha _{1},\ldots,\alpha _{n}\right\}$ with size $k-m(m=1,2)$, such that 
$$a_{21}= a_{11}\sum\limits_{\alpha_{i}\in I_{m}} \alpha_{i},\ \text{or}\  a_{22}= a_{12}\sum\limits_{\alpha_{i}\in I_{m}} \alpha_{i},$$
i.e., Theorem $\ref{ERLAMDS}$ $(2)$ holds.
\iffalse$(ii)$ there exists some subset $J\subseteq\left\{\alpha _{1},\ldots,\alpha _{n}\right\}$ with size $k-2$, such that 
$$a_{21}= a_{11}\sum\limits_{\alpha_{i}\in J} \alpha_{i},\ \text{or}\   a_{22}= a_{12}\sum\limits_{\alpha_{i}\in J} \alpha_{i}.$$
And then we know that the statement (ii) holds if and only if Theorem $\ref{ERLAMDS}$ $(2)$ holds.\fi

From the above, we complete the proof of Theorem  $\ref{ERLAMDS}$.

$\hfill\Box$ 

\iffalse\begin{remark}
For the existence of AMDS $\mathrm{EGRL}_{k,2,0}^{\perp}\left(\boldsymbol{\alpha},\boldsymbol{v},b\right)$, we will prove it by constructing a class of NMDS $\mathrm{EGRL}_{k,2,0}\left(\boldsymbol{\alpha},\boldsymbol{v},b\right)$ in the next section.
\end{remark}\fi
\section{A special construction of NMDS EGRL codes}
 Let $\mathbb{F}_{q}^{*}=\left\{\beta_1,\ldots,\beta_{q-1}\right\}$, in this section,  we construct a class of NMDS codes via EGRL codes, i.e.,  $\mathrm{EGRL}_{k,2,0}\left(\mathbb{F}_{q}^{*},\boldsymbol{1},b\right)$, and then determine its weight distribution for the case  $5\leq k\leq q-2$ and $p=2$,  or $4\leq k\leq q-1$ and $ p\neq 2$. Since their proofs are a little long, for the convenience, we divide the content into the following two subsections.
 
\subsection{The parameters of $\mathrm{EGRL}_{k,2,0}^{\perp}\left(\mathbb{F}_{q}^{*},1,b\right)$}
In this subsection, we prove that when $5\leq k\leq q-2$  and $p=2$, or $4\leq k\leq q-1$ and $p\neq 2$, $\mathrm{EGRL}_{k,2,0}^{\perp}\left(\mathbb{F}_{q}^{*},\boldsymbol{1},b\right)$ is AMDS with the parameters $\left[q+2,q+2-k,k\right]$, and then completely determine its weight distribution.
 
 Firstly, we give the parameters of $\mathrm{EGRL}_{k,2,0}^{\perp}\left(\mathbb{F}_{q}^{*},\boldsymbol{1},b\right)$, especially, the low bound of the minimal distance as the following
\begin{lemma}\label{ERL1dualdistance}
$\mathrm{EGRL}_{k,2,0}^{\perp}\left(\mathbb{F}_{q}^{*},\boldsymbol{1},b\right)$ is an $\left[q+2,q+2-k,d\right]$ linear code with $d\geq k$.
\end{lemma}
{\bf Proof}. By Definition \ref{EGRLdefinition}, it's easy to know that $\mathrm{EGRL}_{k,2,0}^{\perp}\left(\mathbb{F}_{q}^{*},\boldsymbol{1},b\right)$ has the parity-check matrix \begin{equation}\label{ERLgenerator}
	\boldsymbol{G}_{ERL}=\begin{pmatrix}
		1 & \cdots & 1 &0& 0&b \\
		\beta_{1} & \cdots & \beta_{q-1} & 0& 0 & 0 \\
		\vdots & \ddots & \vdots & \vdots & \vdots& \vdots  \\
		\beta_{1}^{k-3} & \cdots & \beta_{q-1}^{k-3} & 0& 0 & 0 \\
		\beta_{1}^{k-2} & \ldots & \beta_{q-1}^{k-2} &  a_{11} &  a_{12}& 0 \\
		\beta_{1}^{k-1} & \cdots & \beta_{q-1}^{k-1} & a_{21} & a_{22}& 0
	\end{pmatrix}_{k\times (n+3)},
\end{equation}
and then  $\mathrm{EGRL}_{k,2,0}^{\perp}\left(\mathbb{F}_{q}^{*},\boldsymbol{1},b\right)$ is an $\left[q+2,q+2-k\right]$ linear code.
%\begin{equation}\label{ERL}
%\boldsymbol{G}_{ERL}=\begin{pmatrix}
%		1 & \cdots & 1 &0& 0&b \\
%		\beta_{1} & \cdots & \beta_{q-1} & 0& 0 & 0 \\
%		\vdots & \ddots & \vdots & \vdots & \vdots& \vdots  \\
%		\beta_{1}^{k-3} & \cdots & \beta_{q-1}^{k-3} & 0& 0 & 0 \\
%		\beta_{1}^{k-2} & \ldots & \beta_{q-1}^{k-2} &  a_{11} &  a_{12}& 0 \\
%		\beta_{1}^{k-1} & \cdots & \beta_{q-1}^{k-1} & a_{21} & a_{22}& 0
%	\end{pmatrix}_{k\times (q+2)}.
%\end{equation}
%where $\boldsymbol{M}_{2\times 2}=\begin{pmatrix}
%	a_{11} & a_{12} \\
%	a_{21} &  a_{22} \\
%\end{pmatrix}\in\mathrm{GL}_{2}(\mathbb{F}_{q}).$

Next, we prove $d\left(\mathrm{EGRL}_{k,2,0}^{\perp}\left(\mathbb{F}_{q}^{*},\boldsymbol{1},b\right)\right)\geq k$. By Corollary 1.4.14 in \cite{A1}, it's enough to prove that any $k-1$ columns of $\boldsymbol{G}_{ERL}$ are $\mathbb{F}_{q}$-linearly
independent, i.e., the submatrix consisted of any $k-1$ columns in $\boldsymbol{G}_{ERL}$ is nonsingular over $\mathbb{F}_{q}$, namely, there exists a $(k-1)\times (k-1)$ non-zero minor for the submatrix consisted of any $k-1$ columns in $\boldsymbol{G}_{ERL}$. In fact, it's easy to know that the submatrix consisted of any $k-1$ columns in $\boldsymbol{G}_{ERL}$  has one of the following forms,
$$\boldsymbol{L}_{1}=\begin{pmatrix}
	1 & \cdots & 1\\
	\beta_{i_1} & \cdots & \beta_{i_{k-1}} \\
	\vdots & \ddots & \vdots \\
	\beta_{i_1}^{k-4} & \cdots & \beta_{i_{k-1}}^{k-4} \\
	\beta_{i_1}^{k-3} & \cdots & \beta_{i_{k-1}}^{k-3} \\
	\beta_{i_1}^{k-2} & \ldots & \beta_{i_{k-1}}^{k-2} \\ 
	\beta_{i_1}^{k-1} & \cdots & \beta_{i_{k-1}}^{k-1}
\end{pmatrix}_{k\times(k-1)},\ \boldsymbol{L}_{2}=\begin{pmatrix}
1 & \cdots & 1&b\\
\beta_{i_1} & \cdots & \beta_{i_{k-2}}&0 \\
\vdots & \ddots & \vdots& \vdots \\
\beta_{i_1}^{k-4} & \cdots & \beta_{i_{k-2}}^{k-4}&0 \\
\beta_{i_1}^{k-3} & \cdots & \beta_{i_{k-2}}^{k-3}&0 \\
\beta_{i_1}^{k-2} & \ldots & \beta_{i_{k-2}}^{k-2}&0 \\ 
\beta_{i_1}^{k-1} & \cdots & \beta_{i_{k-2}}^{k-1}&0
\end{pmatrix}_{k\times(k-1)},$$
$$\boldsymbol{L}_{3,s}=\begin{pmatrix}
	1 & \cdots & 1&0\\
	\beta_{i_1} & \cdots & \beta_{i_{k-2}}&0 \\
	\vdots & \ddots & \vdots& \vdots \\
	\beta_{i_1}^{k-4} & \cdots & \beta_{i_{k-2}}^{k-4}&0 \\
	\beta_{i_1}^{k-3} & \cdots & \beta_{i_{k-2}}^{k-3}&0 \\
	\beta_{i_1}^{k-2} & \ldots & \beta_{i_{k-2}}^{k-2}&a_{1s}\\ 
	\beta_{i_1}^{k-1} & \cdots & \beta_{i_{k-2}}^{k-1}&a_{2s}
\end{pmatrix}_{k\times(k-1)},\ \boldsymbol{L}_{4,s}=\begin{pmatrix}
1 & \cdots & 1&0&b\\
\beta_{i_1} & \cdots & \beta_{i_{k-3}}&0&0 \\
\vdots & \ddots & \vdots& \vdots& \vdots \\
\beta_{i_1}^{k-4} & \cdots & \beta_{i_{k-3}}^{k-4}&0&0 \\
\beta_{i_1}^{k-3} & \cdots & \beta_{i_{k-3}}^{k-3}&0&0 \\
\beta_{i_1}^{k-2} & \ldots & \beta_{i_{k-3}}^{k-2}&a_{1s}&0\\ 
\beta_{i_1}^{k-1} & \cdots & \beta_{i_{k-3}}^{k-1}&a_{2s}&0
\end{pmatrix}_{k\times(k-1)}(s=1,2),$$
$$\boldsymbol{L}_{5}=\begin{pmatrix} 
1 & \cdots & 1&0&0\\
\beta_{i_1} & \cdots & \beta_{i_{k-3}}&0&0 \\
\vdots & \ddots & \vdots& \vdots& \vdots \\
\beta_{i_1}^{k-4} & \cdots & \beta_{i_{k-3}}^{k-4}&0&0 \\
\beta_{i_1}^{k-3} & \cdots & \beta_{i_{k-3}}^{k-3}&0&0 \\
\beta_{i_1}^{k-2} & \ldots & \beta_{i_{k-3}}^{k-2}&a_{11}&a_{12}\\ 
\beta_{i_1}^{k-1} & \cdots & \beta_{i_{k-3}}^{k-1}&a_{21}&a_{22}
\end{pmatrix}_{k\times(k-1)},\ \boldsymbol{L}_{6}=\begin{pmatrix} 
1 & \cdots & 1&0&0&b\\
\beta_{i_1} & \cdots & \beta_{i_{k-4}}&0&0&0 \\
\vdots & \ddots & \vdots& \vdots& \vdots& \vdots \\
\beta_{i_1}^{k-4} & \cdots & \beta_{i_{k-4}}^{k-4}&0&0&0 \\
\beta_{i_1}^{k-3} & \cdots & \beta_{i_{k-4}}^{k-3}&0&0&0 \\
\beta_{i_1}^{k-2} & \ldots & \beta_{i_{k-4}}^{k-2}&a_{11}&a_{12}&0\\ 
\beta_{i_1}^{k-1} & \cdots & \beta_{i_{k-4}}^{k-1}&a_{21}&a_{22}&0
\end{pmatrix}_{k\times(k-1)}.$$ 

It's easy to prove the following statements.

(1) For $\boldsymbol{L}_{1}$ and $\boldsymbol{L}_{5}$, the determinant of the submatrix given by deleting the first row of the corresponding matrix is a $(k-1)\times (k-1)$ non-zero minor.

(2) For $\boldsymbol{L}_{2}$, $\boldsymbol{L}_{3,s}$ and $\boldsymbol{L}_{4,s}$, the determinant of the submatrix given by deleting the last row of the corresponding matrix is a $(k-1)\times (k-1)$ non-zero minor.

(3) For $\boldsymbol{L}_{6}$, the determinant of the submatrix given by deleting the $(k-2)$-th row of the corresponding matrix is a $(k-1)\times (k-1)$ non-zero minor.

From the above discussions, there exists a $(k-1)\times (k-1)$ non-zero minor for the submatrix consisted of any $k-1$ columns in $\boldsymbol{G}_{ERL}$, i.e., the submatrix consisted of any $k-1$ columns in $\boldsymbol{G}_{ERL}$ is nonsingular over $\mathbb{F}_{q}$. 

This completes the proof of Lemma $\ref{ERL1dualdistance}$.

$\hfill\Box$

Secondly, we prove that there does not exist some forms of codewords in $\mathrm{EGRL}_{k,2,0}^{\perp}\left(\boldsymbol{\alpha},\boldsymbol{1},b\right)$ with Hamming weight $k$ as the following 
\begin{lemma}\label{ERLnoexistcodewords} 
For any  $\boldsymbol{c}=(c_{1},\ldots,c_{q-1},c_{q},c_{q+1},c_{q+2})\in\mathrm{EGRL}_{k,2,0}^{\perp}\left(\boldsymbol{\alpha},\boldsymbol{1},b\right)$ with Hamming weight $k$, the following statements are true,

$(1)$ there does not exist any $k$-elements subset $\left\{c_{i_1},\ldots, c_{i_{k}}\right\}$ of $\left\{c_{1},\ldots, c_{q-1}\right\}$ such that  $\prod\limits_{j=1}^{k}c_{i_{j}}\neq 0$ and $c_{q}=c_{q+1}=c_{q+2}=0$;

$(2)$ there does not exist any $(k-1)$-elements subset $\left\{c_{i_1},\ldots, c_{i_{k-1}}\right\}$ of $\left\{c_{1},\ldots, c_{q-1}\right\}$ such that  $\prod\limits_{j=1}^{k-1}c_{i_{j}}\neq 0$ and $c_{q}=c_{q+1}=0$, $c_{q+2}\neq 0$;

$(3)$ there does not exist any $(k-2)$-elements subset $\left\{c_{i_1},\ldots, c_{i_{k-2}}\right\}$ of $\left\{c_{1},\ldots, c_{q-1}\right\}$ such that  $\prod\limits_{j=1}^{k-2}c_{i_{j}}\neq 0$ and $c_{q}c_{q+1}\neq 0$, $c_{q+2}=0$.

$(4)$ there does not exist any $(k-3)$-elements subset $\left\{c_{i_1},\ldots, c_{i_{k-3}}\right\}$ of $\left\{c_{1},\ldots, c_{q-1}\right\}$ such that  $\prod\limits_{j=1}^{k-3}c_{i_{j}}\neq 0$ and $c_{q}c_{q+1}c_{q+2}\neq 0$.
\end{lemma}
{\bf Proof}. (1) It's enough to prove that there does not exist any codeword with Hamming weight $k$ in $\mathrm{EGRL}_{k,2,0}^{\perp}\left(\boldsymbol{\alpha},\boldsymbol{1},b\right)$ as the following form
\begin{equation}\label{nokweightcodeword1}
	(\underbrace{0,\ldots,0,c_{i_1},0,\ldots,0,c_{i_2},\ldots,0,\ldots,0,c_{i_{k-1}},0,\ldots,0,c_{i_k},0,\ldots}_{q-1},0,0,0).
\end{equation}
Otherwise,  if there exists some codeword $\boldsymbol{c}_{11}^{\perp}\in\mathrm{EGRL}_{k,2,0}^{\perp}\left(\boldsymbol{\alpha},\boldsymbol{1},b\right)$ with Hamming weight $k$ as the form $(\ref{nokweightcodeword1})$,
then $\boldsymbol{c}_{11}^{\perp}\cdot \boldsymbol{G}_{ERL}^T=\boldsymbol{0}_{k}$, 
which means that the following homogeneous system of the equations
\begin{equation}\label{lemmaequation1}
	\boldsymbol{N}_{1}\boldsymbol{x}=\mathbf{0}_{k}
\end{equation}
have non-zero solutions, where 
$$\boldsymbol{N}_{1}=
\begin{pmatrix}
	1&\cdots&1\\
	\beta_{i_1}  & \cdots & \beta_{i_{k}}\\
	\vdots&\ddots&\vdots\\
	\beta_{i_1}^{k-3}  & \cdots & \beta_{i_{k}}^{k-3} \\
	\beta_{i_1}^{k-2}  & \cdots & \beta_{i_{k}}^{k-2}\\
	\beta_{i_1}^{k-1} & \cdots & \beta_{i_{k}}^{k-1}  \\
\end{pmatrix}_{k\times k}.
$$
In fact, it's easy to show that
$$\det(\boldsymbol{N}_{1})=\prod_{1\leq j<l\leq k}(\beta_{i_l}-\beta_{i_j})\neq 0,$$	
which means that  $(\ref{lemmaequation1})$ has only zero solution, thus Lemma \ref{ERLnoexistcodewords} (1) is true.

(2) It's enough to prove that there does not exist any codeword with Hamming weight $k$ in $\mathrm{EGRL}_{k,2,0}^{\perp}\left(\boldsymbol{\alpha},\boldsymbol{1},b\right)$ as the following form
\begin{equation}\label{nokweightcodeword2}
	(\underbrace{0,\ldots,0,c_{i_1},0,\ldots,0,c_{i_2},\ldots,0,\ldots,0,c_{i_{k-2}},0,\ldots,0,c_{i_{k-1}},0,\ldots}_{q-1},0,0,c_{i_{k}}),
\end{equation}
Otherwise,  if  there exists some codeword $\boldsymbol{c}_{12}^{\perp}\in\mathrm{EGRL}_{k,2,0}^{\perp}\left(\boldsymbol{\alpha},\boldsymbol{1},b\right)$ with Hamming weight $k$ as the form $(\ref{nokweightcodeword2})$,
then $\boldsymbol{c}_{12}^{\perp}\cdot \boldsymbol{G}_{ERL}^{T}=\boldsymbol{0}_{k}$, 
which means that the following homogeneous system of the equations
\begin{equation}\label{lemmaequation2}
	\boldsymbol{N}_{2}\boldsymbol{x}=\mathbf{0}_{k}
\end{equation}
have non-zero solutions, where 
$$\boldsymbol{N}_{2}=\begin{pmatrix} 
	1&\cdots&1&b\\
	\beta_{i_1} & \cdots & \beta_{i_{k-1}}&0 \\
	\vdots & \ddots & \vdots& \vdots \\
	\beta_{i_1}^{k-3} & \cdots & \beta_{i_{k-1}}^{k-3}&0 \\
	\beta_{i_1}^{k-3} & \ldots & \beta_{i_{k-1}}^{k-2}&0 \\
	\beta_{i_1}^{k-1} & \cdots & \beta_{i_{k-1}}^{k-1}&0 \\
\end{pmatrix}_{k\times k}.
$$
In fact, it's easy to show that 
$$\det(\boldsymbol{N}_{2})=(-1)^{k+1}b\cdot\prod_{t=1}^{k-1}\beta_{i_t}\cdot \prod_{1\leq j<l\leq k-1}(\beta_{i_l}-\beta_{i_j})\neq 0,$$	
it means that $(\ref{lemmaequation2})$ has only zero solution, which is a contradiction. Thus, Lemma \ref{ERLnoexistcodewords} (2)  is true.

(3) It's enough to prove that there does not exist any codeword with Hamming weight $k$ in $\mathrm{EGRL}_{k,2,0}^{\perp}\left(\boldsymbol{\alpha},\boldsymbol{1},b\right)$ as the following form
\begin{equation}\label{nokweightcodeword3}
	(\underbrace{0,\ldots,0,c_{i_1},0,\ldots,0,c_{i_2},\ldots,0,\ldots,0,c_{i_{k-3}},0,\ldots,0,c_{i_{k-2}},0,\ldots}_{q-1},c_{i_{k-1}},c_{i_{k}},0),
\end{equation}
Otherwise,  if  there exists some codeword $\boldsymbol{c}_{13}^{\perp}\in\mathrm{EGRL}_{k,2,0}^{\perp}\left(\boldsymbol{\alpha},\boldsymbol{1},b\right)$ with Hamming weight $k$ as the form $(\ref{nokweightcodeword3})$,
then $\boldsymbol{c}_{13}^{\perp}\cdot \boldsymbol{G}_{ERL}^{T}=\boldsymbol{0}_{k}$, 
which means that the following homogeneous system of the equations
\begin{equation}\label{lemmaequation3}
	\boldsymbol{N}_{3}\boldsymbol{x}=\mathbf{0}_{k}
\end{equation}
have non-zero solutions, where 
$$\boldsymbol{N}_{3}=\begin{pmatrix} 
	1&\cdots&1&0&0\\
	\beta_{i_1} & \cdots & \beta_{i_{k-2}}&0&0 \\
	\vdots & \ddots & \vdots& \vdots& \vdots \\
	\beta_{i_1}^{k-3} & \cdots & \beta_{i_{k-2}}^{k-3}&0&0 \\
	\beta_{i_1}^{k-3} & \ldots & \beta_{i_{k-2}}^{k-2}&a_{11}&a_{12} \\
	\beta_{i_1}^{k-1} & \cdots & \beta_{i_{k-2}}^{k-1}&a_{21}&a_{22} \\
\end{pmatrix}_{k\times k}.
$$
In fact, it's easy to show that 
$$\det(\boldsymbol{N}_{3})=\det(\boldsymbol{M}_{2\times 2})\cdot \prod_{1\leq j<l\leq k-2}(\beta_{i_l}-\beta_{i_j})\neq 0,$$	
it means that $(\ref{lemmaequation3})$ has only zero solution, which is a contradiction. Thus, Lemma \ref{ERLnoexistcodewords} (3)  is true.

(4) It's enough to prove that there does not exist any codeword with Hamming weight $k$ in $\mathrm{EGRL}_{k,2,0}^{\perp}\left(\boldsymbol{\alpha},\boldsymbol{1},b\right)$ as the following form
\begin{equation}\label{nokweightcodeword4}
	(\underbrace{0,\ldots,0,c_{i_1},0,\ldots,0,c_{i_2},\ldots,0,\ldots,0,c_{i_{k-4}},0,\ldots,0,c_{i_{k-3}},0,\ldots}_{q-1},c_{i_{k-2}},c_{i_{k-1}},c_{i_{k}}),
\end{equation}
Otherwise,  if  there exists some codeword $\boldsymbol{c}_{14}^{\perp}\in\mathrm{EGRL}_{k,2,0}^{\perp}\left(\boldsymbol{\alpha},\boldsymbol{1},b\right)$ with Hamming weight $k$ as the form $(\ref{nokweightcodeword4})$,
then $\boldsymbol{c}_{14}^{\perp}\cdot \boldsymbol{G}_{ERL}^{T}=\boldsymbol{0}_{k}$, 
which means that the following homogeneous system of the equations
\begin{equation}\label{lemmaequation4}
	\boldsymbol{N}_{4}\boldsymbol{x}=\mathbf{0}_{k}
\end{equation}
have non-zero solutions, where 
$$\boldsymbol{N}_{4}=\begin{pmatrix} 
	1&\cdots&1&0&0&b\\
	\beta_{i_1} & \cdots & \beta_{i_{k-3}}&0&0&0 \\
	\vdots & \ddots & \vdots& \vdots& \vdots& \vdots \\
	\beta_{i_1}^{k-3} & \cdots & \beta_{i_{k-3}}^{k-3}&0&0&0 \\
	\beta_{i_1}^{k-3} & \ldots & \beta_{i_{k-3}}^{k-2}&a_{11}&a_{12}&0 \\
	\beta_{i_1}^{k-1} & \cdots & \beta_{i_{k-3}}^{k-1}&a_{21}&a_{22}&0 \\
\end{pmatrix}_{k\times k}.
$$
In fact, it's easy to show that 
$$\det(\boldsymbol{N}_{4})=(-1)^{k+1}b\cdot\prod_{t=1}^{k-3}\beta_{i_t}\cdot\det(\boldsymbol{M}_{2\times 2})\cdot \prod_{1\leq j<l\leq k-3}(\beta_{i_l}-\beta_{i_j})\neq 0,$$	
it means that $(\ref{lemmaequation4})$ has only zero solution, which is a contradiction. Thus, Lemma \ref{ERLnoexistcodewords} (4)  is true.

From the above discussions, we complete the proof of Lemma $\ref{ERLnoexistcodewords}$.

 $\hfill\Box$
 
Thirdly, we prove that there exist some forms of codewords in $\mathrm{EGRL}_{k,2,0}^{\perp}\left(\boldsymbol{\alpha},\boldsymbol{1},b\right)$ with Hamming weight $k$, and then determine their numbers as the following 
\begin{lemma}\label{existcodewords}
If $5\leq k\leq q-2$  and $p=2$, or $4\leq k\leq q-1$ and $p\neq 2$, then for any  $\boldsymbol{c}=(c_{1},\ldots,c_{q-1},c_{q},c_{q+1},c_{q+2})\in\mathrm{EGRL}_{k,2,0}^{\perp}\left(\boldsymbol{\alpha},\boldsymbol{1},b\right)$ with Hamming weight $k$, the following statements are true,
	
	$(1)$ if $a_{21}\in\mathbb{F}_{q}$  and $a_{11}\in\mathbb{F}_{q}^{*}$, then the following statements are true,
	
	\ \ \ $(1.1)$ there exist $\# N\left(k-1, \frac{a_{21}}{a_{11}},\mathbb{F}_{q}^{*}\right)$ $(k-1)$-elements subset $\left\{c_{i_1},\ldots, c_{i_{k-1}}\right\}$ of $\left\{c_{1},\ldots, c_{q-1}\right\}$ such that $\prod\limits_{j=1}^{k-1}c_{i_{j}} \neq 0$ and $c_{q} \neq 0, c_{q+1} =c_{q+2}= 0$;

	\ \ \ $(1.2)$ there exist $\# N\left(k-2, \frac{a_{21}}{a_{11}},\mathbb{F}_{q}^{*}\right)$ $(k-2)$-elements subset $\left\{c_{i_1},\ldots, c_{i_{k-2}}\right\}$ of $\left\{c_{1},\ldots, c_{q-1}\right\}$ such that $\prod\limits_{j=1}^{k-2}c_{i_{j}} \neq 0$ and $c_{q}c_{q+2} \neq 0, c_{q+1} = 0$;
	
	$(2)$ if $a_{22}\in\mathbb{F}_{q}$  and $a_{12}\in\mathbb{F}_{q}^{*}$, then the following statements are true,
	
	\ \ \ $(2.1)$ there exist $\# N\left(k-1, \frac{a_{22}}{a_{12}},\mathbb{F}_{q}^{*}\right)$ $(k-1)$-elements subset $\left\{c_{i_1},\ldots, c_{i_{k-1}}\right\}$ of $\left\{c_{1},\ldots, c_{q-1}\right\}$ such that $\prod\limits_{j=1}^{k-1}c_{i_{j}} \neq 0$ and $c_{q}=c_{q+2}=0 , c_{q+1}\neq 0 $;

	\ \ \ $(2.2)$ there exist $\# N\left(k-2, \frac{a_{22}}{a_{12}},\mathbb{F}_{q}^{*}\right)$ $(k-2)$-elements subset $\left\{c_{i_1},\ldots, c_{i_{k-2}}\right\}$ of $\left\{c_{1},\ldots, c_{q-1}\right\}$ such that $\prod\limits_{j=1}^{k-2}c_{i_{j}} \neq 0$ and $c_{q+1}c_{q+2} \neq 0, c_{q} = 0$;
	
	$(3)$ if $a_{21}\in\mathbb{F}_{q}^{*}$ and $a_{11}=0$, then the following statements are true,
	
	\ \ \ $(3.1)$ there does not exist any  $(k-1)$-elements subset $\left\{c_{i_1},\ldots, c_{i_{k-1}}\right\}$ of $\left\{c_{1},\ldots, c_{q-1}\right\}$ such that $\prod\limits_{j=1}^{k-1}c_{i_{j}} \neq 0$ and $c_{q} \neq 0, c_{q+1} =c_{q+2}= 0$;

	\ \ \ $(3.2)$ there does not exist any  $(k-2)$-elements subset $\left\{c_{i_1},\ldots, c_{i_{k-2}}\right\}$ of $\left\{c_{1},\ldots, c_{q-1}\right\}$ such that $\prod\limits_{j=1}^{k-2}c_{i_{j}} \neq 0$ and $c_{q}c_{q+2} \neq 0, c_{q+1} = 0$;
	
	$(4)$ if $a_{22}\in\mathbb{F}_{q}^{*}$ and $a_{12}=0$, then the following statements are true,
	
	\ \ \ $(4.1)$ there does not exist any $(k-1)$-elements subset $\left\{c_{i_1},\ldots, c_{i_{k-1}}\right\}$ of $\left\{c_{1},\ldots, c_{q-1}\right\}$ such that $\prod\limits_{j=1}^{k-1}c_{i_{j}} \neq 0$ and $c_{q+1} \neq 0, c_{q} =c_{q+2}= 0$;
 
	\ \ \ $(4.2)$ there does not exist any  $(k-2)$-elements subset $\left\{c_{i_1},\ldots, c_{i_{k-2}}\right\}$ of $\left\{c_{1},\ldots, c_{q-1}\right\}$ such that $\prod\limits_{j=1}^{k-2}c_{i_{j}} \neq 0$ and $c_{q+1}c_{q+2} \neq 0, c_{q} = 0$. 
\end{lemma}	
{\bf Proof}. (1) Firstly, we prove (1.1), it's enough to prove that there exists some codeword with Hamming weight $k$ in $\mathrm{EGRL}_{k,2,0}^{\perp}\left(\boldsymbol{\alpha},\boldsymbol{1},b\right)$ as the following form
$$(\underbrace{0,\ldots,0,c_{i_1},0,\ldots,0,c_{i_2},\ldots,0,\ldots,0,c_{i_{k-2}},0,\ldots,0,c_{i_{k-1}},0,\ldots}_{q-1},c_{i_k},0,0),$$
i.e., we only need to prove that the following homogeneous system of the equations
\begin{equation}\label{kweightsystem1}
	\boldsymbol{Q}_{1}\boldsymbol{x}=\mathbf{0}_{k}
\end{equation}
have non-zero solutions, where 
$$\boldsymbol{Q}_{1}=
\begin{pmatrix}
	1&\cdots&1&0\\
	\beta_{i_1}  & \cdots & \beta_{i_{k-1}}& 0  \\
	\vdots&\ddots&\vdots&\vdots\\
	\beta_{i_1}^{k-3}  & \cdots & \beta_{i_{k-1}}^{k-3}&0  \\
	\beta_{i_1}^{k-2}  & \cdots & \beta_{i_{k-1}}^{k-2}& a_{11}  \\
	\beta_{i_1}^{k-1} & \cdots & \beta_{i_{k-1}}^{k-1}& a_{21}  \\
\end{pmatrix}_{k\times k}.$$
In fact, by Lemma \ref{Vandermonde}, we have
$$\det( \boldsymbol{Q}_{1}) =\left(a_{21}-a_{11}\sum\limits_{t=1}^{k-1} \beta_{i_t}\right)\cdot\prod\limits_{\substack{1 \leq j < l \leq k-1}} (\beta_{i_l} - \beta_{i_j}).$$ 
Note that $\prod\limits_{1 \leq j< l\leq k-2}\left(\beta_{i_{l}}-\beta_{i_{j}}\right)\neq 0$, then it's enough to prove that there exists some $(k-1)$-elements subset $\left\{\beta_{i_1},\ldots, \beta_{i_{k-1}}\right\}$ of $\mathbb{F}_{q}^{*}$ such that $a_{21}-a_{11}\sum\limits_{t=1}^{k-1} \beta_{i_t}=0$. Furthermore, by Lemmas $\ref{subsetsumneq0}$-$\ref{subsetsum0}$, we have $\# N\left(k-1, \frac{a_{21}}{a_{11}},\mathbb{F}_{q}^{*}\right)\neq 0$ for $a_{21}\in\mathbb{F}_{q}$ and $a_{11}\in\mathbb{F}_{q}^{*}$, which means that there exist $\# N\left(k-1, \frac{a_{21}}{a_{11}},\mathbb{F}_{q}^{*}\right)$  $(k-1)$-elements subset $\left\{\beta_{i_1},\ldots, \beta_{i_{k-1}}\right\}$ of $\mathbb{F}_{q}^{*}$ such that $\sum\limits_{t=1}^{k-1} \beta_{i_t}= \frac{a_{21}}{a_{11}}$, i.e., $a_{21}-a_{11}\sum\limits_{t=1}^{k-1} \beta_{i_t}=0$, thus we prove (1.1).

Now for (1.2), it's enough to prove that there exists some codeword with Hamming weight $k$ in $\mathrm{EGRL}_{k,2,0}^{\perp}\left(\boldsymbol{\alpha},\boldsymbol{1},b\right)$ as the following form
$$(\underbrace{0,\ldots,0,c_{i_1},0,\ldots,0,c_{i_2},\ldots,0,\ldots,0,c_{i_{k-3}},0,\ldots,0,c_{i_{k-2}},0,\ldots}_{q-1},c_{i_{k-1}},0,c_{i_k}),$$
i.e., we only need to prove that the following homogeneous system of the equations
\begin{equation}\label{kweightsystem2}
	\boldsymbol{Q}_{2}\boldsymbol{x}=\mathbf{0}_{k}
\end{equation}
have non-zero solutions, where 
$$\boldsymbol{Q}_{2}=
\begin{pmatrix}
1&\cdots&1&0&b\\
\beta_{i_1}  & \cdots & \beta_{i_{k-2}}& 0&0  \\
\vdots&\ddots&\vdots&\vdots&\vdots\\
\beta_{i_1}^{k-3}  & \cdots & \beta_{i_{k-2}}^{k-3}&0&0  \\
\beta_{i_1}^{k-2}  & \cdots & \beta_{i_{k-2}}^{k-2}& a_{11}&0  \\
\beta_{i_1}^{k-1} & \cdots & \beta_{i_{k-2}}^{k-1}& a_{21}&0  \\
\end{pmatrix}_{k\times k}.
$$
In fact, by Lemma \ref{Vandermonde}, we have
$$\det( \boldsymbol{Q}_{2}) =b\prod_{t=1}^{k-2}\beta_{i_t}\cdot\left(a_{21}-a_{11}\sum\limits_{t=1}^{k-2} \beta_{i_t}\right)\cdot\prod\limits_{\substack{1 \leq j < l \leq k-2}} (\beta_{i_l} - \beta_{i_j}).$$
Note that $\prod\limits_{1 \leq j< l\leq k-2}\left(\beta_{i_{l}}-\beta_{i_{j}}\right)\neq 0$, then it's enough to prove that there exists some $(k-2)$-elements subset $\left\{\beta_{i_1},\ldots, \beta_{i_{k-2}}\right\}$ of $\mathbb{F}_{q}^{*}$ such that $a_{21}-a_{11}\sum\limits_{t=1}^{k-2} \beta_{i_t}=0$. Furthermore, by Lemmas $\ref{subsetsumneq0}$-$\ref{subsetsum0}$, we have $\# N\left(k-2, \frac{a_{21}}{a_{11}},\mathbb{F}_{q}^{*}\right)\neq 0$ for $a_{21}\in\mathbb{F}_{q}$ and $a_{11}\in\mathbb{F}_{q}^{*}$, which means that there exist $\# N\left(k-2, \frac{a_{21}}{a_{11}},\mathbb{F}_{q}^{*}\right)$  $(k-2)$-elements subset $\left\{\beta_{i_1},\ldots, \beta_{i_{k-2}}\right\}$ of $\mathbb{F}_{q}^{*}$ such that $\sum\limits_{t=1}^{k-2} \beta_{i_t}= \frac{a_{21}}{a_{11}}$, i.e., $a_{21}-a_{11}\sum\limits_{t=1}^{k-2} \beta_{i_t}=0$, thus we prove (1.2).

(2) In the similar proofs as those of (1) (1.1)-(1.2), we immediately get (2.1)-(2.2), respectively.

(3) It's enough to prove that there does not exists  any codeword with Hamming weight $k$ in $\mathrm{EGRL}_{k,2,0}^{\perp}\left(\boldsymbol{\alpha},\boldsymbol{1},b\right)$ which has one of the following forms,
$$(\underbrace{0,\ldots,0,c_{i_1},0,\ldots,0,c_{i_2},\ldots,0,\ldots,0,c_{i_{k-2}},0,\ldots,0,c_{i_{k-1}},0,\ldots}_{q-1},c_{i_k},0,0),$$ 
$$(\underbrace{0,\ldots,0,c_{i_1},0,\ldots,0,c_{i_2},\ldots,0,\ldots,0,c_{i_{k-3}},0,\ldots,0,c_{i_{k-2}},0,\ldots}_{q-1},c_{i_{k-1}},0,c_{i_k}),$$
i.e., we only need to prove that both the following two homogeneous systems of the equations 
\begin{equation}\label{kweightsystem3}
	\boldsymbol{Q}_{3}\boldsymbol{x}=\mathbf{0}_{k}
\end{equation}
and 
\begin{equation}\label{kweightsystem4}
	\boldsymbol{Q}_{4}\boldsymbol{x}=\mathbf{0}_{k}
\end{equation}
have only zero solution, where 
$$\boldsymbol{Q}_{3}=
\begin{pmatrix}
	1&\cdots&1&0\\
	\beta_{i_1}  & \cdots & \beta_{i_{k-1}}& 0  \\
	\vdots&\ddots&\vdots&\vdots\\
	\beta_{i_1}^{k-3}  & \cdots & \beta_{i_{k-1}}^{k-3}&0  \\
	\beta_{i_1}^{k-2}  & \cdots & \beta_{i_{k-1}}^{k-2}& 0 \\
	\beta_{i_1}^{k-1} & \cdots & \beta_{i_{k-1}}^{k-1}& a_{21}  \\
\end{pmatrix}_{k\times k},\boldsymbol{Q}_{4s}=
\begin{pmatrix}
1&\cdots&1&0&b\\
\beta_{i_1}  & \cdots & \beta_{i_{k-2}}& 0&0  \\
\vdots&\ddots&\vdots&\vdots&\vdots\\
\beta_{i_1}^{k-3}  & \cdots & \beta_{i_{k-2}}^{k-3}&0 &0 \\
\beta_{i_1}^{k-2}  & \cdots & \beta_{i_{k-2}}^{k-2}& 0&0 \\
\beta_{i_1}^{k-1} & \cdots & \beta_{i_{k-2}}^{k-1}& a_{21}&0  \\
\end{pmatrix}_{k\times k}.$$
While by Lemma \ref{Vandermonde}, we have
$$\det( \boldsymbol{Q}_{3}) =a_{21}\prod\limits_{\substack{1 \leq j < l \leq k-1}} (\beta_{i_l} - \beta_{i_j})\neq 0$$
and
$$\det( \boldsymbol{Q}_{4}) =(-1)^{k+1}b\cdot a_{21}\cdot\prod\limits_{\substack{1 \leq j < l \leq k-2}} (\beta_{i_l} - \beta_{i_j})\neq 0,$$
which means that both $(\ref{lemmaequation3})$ and $(\ref{lemmaequation4})$ have only zero solution, thus (3) is true.	

(4) In the similar proofs as those of (3) (3.1)-(3.2), we immediately get (4.1)-(4.2), respectively.

From the above discussions, we complete the proof of Lemma $\ref{existcodewords}$.

$\hfill\Box$

Finally, based on the above Lemmas $\ref{ERL1dualdistance}$-$\ref{existcodewords}$, we prove that when $5\leq k\leq q-2$  and $p=2$, or $4\leq k\leq q-1$ and $p\neq 2$,  $\mathrm{EGRL}_{k,2,0}^{\perp}\left(\mathbb{F}_{q}^{*},\boldsymbol{1},b\right)$ is AMDS with the parameters $\left[q+2,q+2-k,k\right]$,  and obtain the number of codewords with weight $k$ in $\mathrm{EGRL}_{k,2,0}^{\perp}\left(\boldsymbol{\alpha},\boldsymbol{1},b\right)$ as the following
\begin{theorem}\label{ERLdualweight}
	For $5\leq k\leq q-2$  and $p=2$, or $4\leq k\leq q-1$ and $p\neq 2$, $\mathrm{EGRL}_{k,2,0}^{\perp}\left(\mathbb{F}_{q}^{*},\boldsymbol{1},b\right)$ is AMDS with the parameters $\left[q+2,q+2-k,k\right]$. And the total number $A_k^{\bot}$ of codewords with weight $k$ in $\mathrm{EGRL}_{k,2,0}^{\perp}\left(\boldsymbol{\alpha},\boldsymbol{1},b\right)$ is
	$$\scriptsize A_k^{\bot}=\begin{cases}
		 \frac{2(q-1)}{q}\binom{q}{k-1}+(-1)^{k-1}\frac{2(q-1)}{q}\left((-1)^{\lfloor\frac{k-1}{p}\rfloor+1}\binom{\frac{q}{p}-1}{\lfloor\frac{k-1}{p}\rfloor}+(-1)^{\lfloor\frac{k-2}{p}\rfloor}\binom{\frac{q}{p}-1}{\lfloor\frac{k-2}{p}\rfloor}\right),&\text{if~}a_{11} a_{21} a_{12} a_{22}\neq 0;\\
		 \frac{2(q-1)}{q}\binom{q}{k-1}+(-1)^{k-2}\frac{(q-1)(q-2)}{q}\left((-1)^{\lfloor\frac{k-1}{p}\rfloor+1}\binom{\frac{q}{p}-1}{\lfloor\frac{k-1}{p}\rfloor}+ (-1)^{\lfloor\frac{k-2}{p}\rfloor}\binom{\frac{q}{p}-1}{\lfloor\frac{k-2}{p}\rfloor}\right),&\text{if~}a_{11} a_{12}a_{21}\neq 0\ \text{and~}a_{22}= 0;\\
		 &\ \ \ \text{or~} a_{11}a_{12}a_{22}\neq 0\ \text{and~}\ a_{21}=0;\\
		 \frac{q-1}{q}\binom{q}{k-1}+(-1)^{k-1}\frac{q-1}{q}\left((-1)^{\lfloor\frac{k-1}{p}\rfloor+1}\binom{\frac{q}{p}-1}{\lfloor\frac{k-1}{p}\rfloor}+(-1)^{\lfloor\frac{k-2}{p}\rfloor}\binom{\frac{q}{p}-1}{\lfloor\frac{k-2}{p}\rfloor}\right),&\text{if~}a_{11} a_{21}a_{12}\neq 0\ \text{and~}a_{12}=0;\\
		 &\ \ \ \text{or~}a_{12}a_{21}a_{22}\neq 0\ \text{and~}a_{11}= 0;\\
		%\frac{(q-1)}{q}\binom{q}{k-1}+(-1)^{k-3}\frac{q-1}{q}\left((-1)^{\lfloor\frac{k-1}{p}\rfloor+1}\binom{\frac{q}{p}-1}{\lfloor\frac{k-1}{p}\rfloor}+(-1)^{\lfloor\frac{k-2}{p}\rfloor}\binom{\frac{q}{p}-1}{\lfloor\frac{k-2}{p}\rfloor} \right),
		 \frac{(q-1)}{q}\binom{q}{k-1}+(-1)^{k-2}\frac{(q-1)^2}{q}\left((-1)^{\lfloor\frac{k-1}{p}\rfloor+1}\binom{\frac{q}{p}-1}{\lfloor\frac{k-1}{p}\rfloor}+(-1)^{\lfloor\frac{k-2}{p}\rfloor}\binom{\frac{q}{p}-1}{\lfloor\frac{k-2}{p}\rfloor} \right),&\text{if~}a_{11}a_{22}\neq 0\ \text{and~}a_{12}=a_{21}=0;\\
		 &\ \ \ \text{or~} a_{12}a_{21}\neq 0\ \text{and~} a_{11}=a_{22}=0.\\
		 %\frac{2(q-1)}{q}\binom{q}{k-1}+(-1)^{k-2}\frac{(q-1)(q-2)}{q}\left((-1)^{\lfloor\frac{k-1}{p}\rfloor+1}\binom{\frac{q}{p}-1}{\lfloor\frac{k-1}{p}\rfloor}+(-1)^{\lfloor\frac{k-2}{p}\rfloor}\binom{\frac{q}{p}-1}{\lfloor\frac{k-2}{p}\rfloor} \right),
		 %\frac{(q-1)}{q}\binom{q}{k-1}+(-1)^{k-2}\frac{(q-1)^2}{q}\left((-1)^{\lfloor\frac{k-1}{p}\rfloor+1}\binom{\frac{q}{p}-1}{\lfloor\frac{k-1}{p}\rfloor}+(-1)^{\lfloor\frac{k-2}{p}\rfloor}\binom{\frac{q}{p}-1}{\lfloor\frac{k-2}{p}\rfloor} \right),
	\end{cases}$$
\end{theorem}
{\bf Proof}. It's easy to prove that $\rank(\boldsymbol{G}_{ERL})=k$, hence, $${\rm dim}(\mathrm{EGRL}_{k,2,0}\left(\mathbb{F}_{q}^{*},\boldsymbol{1},b\right))=\rank(\boldsymbol{G}_{ERL})=k,$$ which means that $$\dim{(\mathrm{EGRL}_{k,2,0}^{\perp}\left(\boldsymbol{\alpha},\boldsymbol{1},b\right)}=q+2-\dim{(\mathrm{EGRL}_{k,2,0}\left(\mathbb{F}_{q}^{*},\boldsymbol{1},b\right))}=q+2-k.$$ In addition, by Lemmas $\ref{ERL1dualdistance}$-$\ref{existcodewords}$, we have $d\left(\mathrm{EGRL}_{k,2,0}^{\perp}\left(\boldsymbol{\alpha},\boldsymbol{1},b\right)\right)\geq k$ and when $5\leq k\leq q-2$  and $p=2$, or $4\leq k\leq q-1$ and $p\neq 2$, there exists some codeword $\boldsymbol{c}\in\mathrm{EGRL}_{k,2,0}^{\perp}\left(\boldsymbol{\alpha},\boldsymbol{1},b\right)$ with Hamming weight $k$, and then $$d\left(\mathrm{EGRL}_{k,2,0}^{\perp}\left(\boldsymbol{\alpha},\boldsymbol{1},b\right)\right)=k.$$ 
Furthermore, for $5\leq k\leq q-2$  and $p=2$, or $4\leq k\leq q-1$ and $p\neq 2$, $\mathrm{EGRL}_{k,2,0}^{\perp}\left(\mathbb{F}_{q}^{*},\boldsymbol{1},b\right)$ is AMDS with the parameters $\left[q+2,q+2-k,k\right]$.

Next, we calculate the number of codewords with weight $k$ in $\mathrm{EGRL}_{k,2,0}^{\perp}\left(\boldsymbol{\alpha},\boldsymbol{1},b\right)$ as the following seven cases.

\textbf{Case 1.} If $a_{11}a_{21}a_{12}a_{22}\neq 0$, then by the proofs of Lemma \ref{existcodewords} (1)-(2), we know that for any given $(k-1)$-elements subset or $(k-2)$-elements subset, the corresponding homogeneous systems of the equations $( \ref{kweightsystem1})$-$( \ref{kweightsystem2})$ have $(q-1)$ non-zero solutions, respectively. It means that the total number of codewords with weight $k$ in $\mathrm{EGRL}_{k,2,0}^{\perp}\left(\boldsymbol{\alpha},\boldsymbol{1},b\right)$ is  
$$\begin{aligned}
	A_k^{\bot}=&(q-1)\# N\left(k-1, \frac{a_{21}}{a_{11}},\mathbb{F}_{q}^{*}\right)+(q-1)\# N\left(k-2, \frac{a_{21}}{a_{11}},\mathbb{F}_{q}^{*}\right)\\
	&+(q-1)\# N\left(k-1, \frac{a_{22}}{a_{12}},\mathbb{F}_{q}^{*}\right)+(q-1)\# N\left(k-2, \frac{a_{22}}{a_{12}},\mathbb{F}_{q}^{*}\right)\\
	%=&\frac{2(q-1)}{q}{q-1\choose k-1}+(-1)^{k-2+\lfloor\frac{k-1}{p}\rfloor}\frac{2(q-1)}{q}{\frac{q}{p}-1\choose\lfloor\frac{k-1}{p}\rfloor}\\
	%&+\frac{2(q-1)}{q}\binom{q-1}{k-2}+(-1)^{k-3+\lfloor\frac{k-2}{p}\rfloor}\frac{2(q-1)}{q}\binom{\frac{q}{p}-1}{\lfloor\frac{k-2}{p}\rfloor}\\
	=&\frac{2(q-1)}{q}\binom{q}{k-1}+(-1)^{k-1}\frac{2(q-1)}{q}\left((-1)^{\lfloor\frac{k-1}{p}\rfloor+1}\binom{\frac{q}{p}-1}{\lfloor\frac{k-1}{p}\rfloor}+(-1)^{\lfloor\frac{k-2}{p}\rfloor}\binom{\frac{q}{p}-1}{\lfloor\frac{k-2}{p}\rfloor}\right).
\end{aligned}$$	

\textbf{Case 2.} if $a_{11}a_{12}a_{21}\neq 0$ and $a_{22}=0$, then by Lemma \ref{existcodewords} (1)-(2), the total number of codewords with weight $k$ in $\mathrm{EGRL}_{k,2,0}^{\perp}\left(\boldsymbol{\alpha},\boldsymbol{1},b\right)$ is
$$\begin{aligned}
	A_k^{\bot}=&(q-1)\# N\left(k-1, \frac{a_{21}}{a_{11}},\mathbb{F}_{q}^{*}\right)+(q-1)\# N\left(k-2, \frac{a_{21}}{a_{11}},\mathbb{F}_{q}^{*}\right)\\
	&+(q-1)\# N\left(k-1,0,\mathbb{F}_{q}^{*}\right)+(q-1)\# N\left(k-2,0,\mathbb{F}_{q}^{*}\right)\\
	%=&\frac{2(q-1)}{q}\binom{q}{k-1}+(-1)^{k-1}\frac{(q-1)}{q}\left((-1)^{\lfloor\frac{k-1}{p}\rfloor+1}\binom{\frac{q}{p}-1}{\lfloor\frac{k-1}{p}\rfloor}+ (-1)^{\lfloor\frac{k-2}{p}\rfloor}\binom{\frac{q}{p}-1}{\lfloor\frac{k-2}{p}\rfloor}\right)\\
	%&+(-1)^{k-2}\frac{(q-1)^2}{q}\left((-1)^{\lfloor\frac{k-1}{p}\rfloor+1}\binom{\frac{q}{p}-1}{\lfloor\frac{k-1}{p}\rfloor}+(-1)^{\lfloor\frac{k-2}{p}\rfloor}\binom{\frac{q}{p}-1}{\lfloor\frac{k-2}{p}\rfloor}\right)\\
	=&\frac{2(q-1)}{q}\binom{q}{k-1}+(-1)^{k-2}\frac{(q-1)(q-2)}{q}\left((-1)^{\lfloor\frac{k-1}{p}\rfloor+1}\binom{\frac{q}{p}-1}{\lfloor\frac{k-1}{p}\rfloor}+ (-1)^{\lfloor\frac{k-2}{p}\rfloor}\binom{\frac{q}{p}-1}{\lfloor\frac{k-2}{p}\rfloor}\right).
\end{aligned}$$	

\textbf{Case 3.} if $a_{11}a_{21}a_{22}\neq 0$ and $a_{12}=0$, then by Lemma \ref{existcodewords} (1) and Lemma \ref{existcodewords} (4),  the total number of codewords with weight $k$ in $\mathrm{EGRL}_{k,2,0}^{\perp}\left(\boldsymbol{\alpha},\boldsymbol{1},b\right)$ is
$$\begin{aligned}
	A_k^{\bot}=&(q-1)\# N\left(k-1, \frac{a_{21}}{a_{11}},\mathbb{F}_{q}^{*}\right)+(q-1)\# N\left(k-2, \frac{a_{21}}{a_{11}},\mathbb{F}_{q}^{*}\right)\\
	=&\frac{q-1}{q}\binom{q}{k-1}+(-1)^{k-1}\frac{q-1}{q}\left((-1)^{\lfloor\frac{k-1}{p}\rfloor+1}\binom{\frac{q}{p}-1}{\lfloor\frac{k-1}{p}\rfloor}+(-1)^{\lfloor\frac{k-2}{p}\rfloor}\binom{\frac{q}{p}-1}{\lfloor\frac{k-2}{p}\rfloor}\right).
\end{aligned}$$	

\textbf{Case 4.} if $a_{11}a_{12}a_{22}\neq0$ and $a_{21}=0$, then by Lemma \ref{existcodewords} (1)-(2),  the total number of codewords with weight $k$ in $\mathrm{EGRL}_{k,2,0}^{\perp}\left(\boldsymbol{\alpha},\boldsymbol{1},b\right)$ is 
$$\begin{aligned}
	A_k^{\bot}=&(q-1)\# N\left(k-1,0,\mathbb{F}_{q}^{*}\right)+(q-1)\# N\left(k-2,0,\mathbb{F}_{q}^{*}\right)\\
	&+(q-1)\# N\left(k-1, \frac{a_{22}}{a_{12}},\mathbb{F}_{q}^{*}\right)+(q-1)\# N\left(k-2, \frac{a_{22}}{a_{12}},\mathbb{F}_{q}^{*}\right)\\
	%=&\frac{2(q-1)}{q}\binom{q}{k-1}+(-1)^{k-1}\frac{(q-1)^2}{q}\left((-1)^{\lfloor\frac{k-1}{p}\rfloor+1}\binom{\frac{q}{p}-1}{\lfloor\frac{k-1}{p}\rfloor}+(-1)^{\lfloor\frac{k-2}{p}\rfloor}\binom{\frac{q}{p}-1}{\lfloor\frac{k-2}{p}\rfloor} \right)\\
	%+&(-1)^{k-2}\frac{(q-1)}{q}\left((-1)^{\lfloor\frac{k-1}{p}\rfloor+1}\binom{\frac{q}{p}-1}{\lfloor\frac{k-1}{p}\rfloor}+(-1)^{\lfloor\frac{k-2}{p}\rfloor}\binom{\frac{q}{p}-1}{\lfloor\frac{k-2}{p}\rfloor} \right)\\
	=&\frac{2(q-1)}{q}\binom{q}{k-1}+(-1)^{k-2}\frac{(q-1)(q-2)}{q}\left((-1)^{\lfloor\frac{k-1}{p}\rfloor+1}\binom{\frac{q}{p}-1}{\lfloor\frac{k-1}{p}\rfloor}+(-1)^{\lfloor\frac{k-2}{p}\rfloor}\binom{\frac{q}{p}-1}{\lfloor\frac{k-2}{p}\rfloor} \right).
\end{aligned}$$

\textbf{Case 5.} if $a_{11}a_{22}\neq 0$ and $a_{12}=a_{21}=0$, then by Lemma \ref{existcodewords} (1) and Lemma \ref{existcodewords} (4),  the total number of codewords with weight $k$ in $\mathrm{EGRL}_{k,2,0}^{\perp}\left(\boldsymbol{\alpha},\boldsymbol{1},b\right)$ is 
$$\begin{aligned}
	A_k^{\bot}=&(q-1)\# N\left(k-1,0,\mathbb{F}_{q}^{*}\right)+(q-1)\# N\left(k-2,0,\mathbb{F}_{q}^{*}\right)\\
	=&\frac{(q-1)}{q}\binom{q}{k-1}+(-1)^{k-2}\frac{(q-1)^2}{q}\left((-1)^{\lfloor\frac{k-1}{p}\rfloor+1}\binom{\frac{q}{p}-1}{\lfloor\frac{k-1}{p}\rfloor}+(-1)^{\lfloor\frac{k-2}{p}\rfloor}\binom{\frac{q}{p}-1}{\lfloor\frac{k-2}{p}\rfloor} \right).
\end{aligned}$$	

\textbf{Case 6.} if $a_{12}a_{21}a_{22}\neq 0$ and $a_{11}=0$, then by Lemma \ref{existcodewords} (2)-(3),  the total number of codewords with weight $k$ in $\mathrm{EGRL}_{k,2,0}^{\perp}\left(\boldsymbol{\alpha},\boldsymbol{1},b\right)$ is 
$$\begin{aligned}
	A_k^{\bot}=&(q-1)\# N\left(k-1, \frac{a_{22}}{a_{12}},\mathbb{F}_{q}^{*}\right)+(q-1)\# N\left(k-2, \frac{a_{22}}{a_{12}},\mathbb{F}_{q}^{*}\right)\\
	=&\frac{(q-1)}{q}\binom{q}{k-1}+(-1)^{k-1}\frac{q-1}{q}\left((-1)^{\lfloor\frac{k-1}{p}\rfloor+1}\binom{\frac{q}{p}-1}{\lfloor\frac{k-1}{p}\rfloor}+(-1)^{\lfloor\frac{k-2}{p}\rfloor}\binom{\frac{q}{p}-1}{\lfloor\frac{k-2}{p}\rfloor} \right).
\end{aligned}$$	

\textbf{Case 7.} if $a_{12}a_{21}\neq 0$ and $a_{11}=a_{22}= 0$, then by Lemma \ref{existcodewords} (2)-(3),   the total number of codewords with weight $k$ in $\mathrm{EGRL}_{k,2,0}^{\perp}\left(\boldsymbol{\alpha},\boldsymbol{1},b\right)$ is 
$$\begin{aligned}
	A_k^{\bot}=&(q-1)\# N\left(k-1,0,\mathbb{F}_{q}^{*}\right)+(q-1)\# N\left(k-2,0,\mathbb{F}_{q}^{*}\right)\\
	=&\frac{(q-1)}{q}\binom{q}{k-1}+(-1)^{k-2}\frac{(q-1)^2}{q}\left((-1)^{\lfloor\frac{k-1}{p}\rfloor+1}\binom{\frac{q}{p}-1}{\lfloor\frac{k-1}{p}\rfloor}+(-1)^{\lfloor\frac{k-2}{p}\rfloor}\binom{\frac{q}{p}-1}{\lfloor\frac{k-2}{p}\rfloor} \right).
\end{aligned}$$	

From the above discussions, we complete the proof of Theorem $\ref{ERLdualweight}$.

$\hfill\Box$

\subsection{The parameters of $\mathrm{EGRL}_{k,2,0} \left(\mathbb{F}_{q}^{*},1,b\right)$}
In this subsection, we prove that when $5\leq k\leq q-2$  and $p=2$, or $4\leq k\leq q-1$ and $p\neq 2$, $\mathrm{EGRL}_{k,2,0}\left(\mathbb{F}_{q}^{*},\boldsymbol{1},b\right)$ is NMDS and completely determine its weight distribution. 
\begin{theorem}\label{ERL1}
For $5\leq k\leq q-2$  and $p=2$, or $4\leq k\leq q-1$ and $p\neq 2$, $\mathrm{EGRL}_{k,2,0}\left(\mathbb{F}_{q}^{*},\boldsymbol{1},b\right)$ is NMDS with the parameters $[q+2,k,q+2-k]_q$, and the total number  $A_{q+2-k}^{\bot}$  of codewords with weight $q+2-k$ in $\mathrm{EGRL}_{k,2,0}\left(\mathbb{F}_{q}^{*},\boldsymbol{1},b\right)$ is
	$$\scriptsize A_{q+2-k}=\begin{cases}
		\frac{2(q-1)}{q}\binom{q}{k-1}+(-1)^{k-1}\frac{2(q-1)}{q}\left((-1)^{\lfloor\frac{k-1}{p}\rfloor+1}\binom{\frac{q}{p}-1}{\lfloor\frac{k-1}{p}\rfloor}+(-1)^{\lfloor\frac{k-2}{p}\rfloor}\binom{\frac{q}{p}-1}{\lfloor\frac{k-2}{p}\rfloor}\right),&\text{if~}a_{11} a_{21} a_{12} a_{22}\neq 0;\\
		\frac{2(q-1)}{q}\binom{q}{k-1}+(-1)^{k-2}\frac{(q-1)(q-2)}{q}\left((-1)^{\lfloor\frac{k-1}{p}\rfloor+1}\binom{\frac{q}{p}-1}{\lfloor\frac{k-1}{p}\rfloor}+ (-1)^{\lfloor\frac{k-2}{p}\rfloor}\binom{\frac{q}{p}-1}{\lfloor\frac{k-2}{p}\rfloor}\right),&\text{if~}a_{11} a_{12}a_{21}\neq 0\ \text{and~}a_{22}= 0;\\
		&\ \ \ \text{or~} a_{11}a_{12}a_{22}\neq 0\ \text{and~}\ a_{21}=0;\\
		\frac{q-1}{q}\binom{q}{k-1}+(-1)^{k-1}\frac{q-1}{q}\left((-1)^{\lfloor\frac{k-1}{p}\rfloor+1}\binom{\frac{q}{p}-1}{\lfloor\frac{k-1}{p}\rfloor}+(-1)^{\lfloor\frac{k-2}{p}\rfloor}\binom{\frac{q}{p}-1}{\lfloor\frac{k-2}{p}\rfloor}\right),&\text{if~}a_{11} a_{21}a_{12}\neq 0\ \text{and~}a_{12}=0;\\
		&\ \ \ \text{or~}a_{12}a_{21}a_{22}\neq 0\ \text{and~}a_{11}= 0;\\
		%\frac{(q-1)}{q}\binom{q}{k-1}+(-1)^{k-3}\frac{q-1}{q}\left((-1)^{\lfloor\frac{k-1}{p}\rfloor+1}\binom{\frac{q}{p}-1}{\lfloor\frac{k-1}{p}\rfloor}+(-1)^{\lfloor\frac{k-2}{p}\rfloor}\binom{\frac{q}{p}-1}{\lfloor\frac{k-2}{p}\rfloor} \right),
		\frac{(q-1)}{q}\binom{q}{k-1}+(-1)^{k-2}\frac{(q-1)^2}{q}\left((-1)^{\lfloor\frac{k-1}{p}\rfloor+1}\binom{\frac{q}{p}-1}{\lfloor\frac{k-1}{p}\rfloor}+(-1)^{\lfloor\frac{k-2}{p}\rfloor}\binom{\frac{q}{p}-1}{\lfloor\frac{k-2}{p}\rfloor} \right),&\text{if~}a_{11}a_{22}\neq 0\ \text{and~}a_{12}=a_{21}=0;\\
		&\ \ \ \text{or~} a_{12}a_{21}\neq 0\ \text{and~} a_{11}=a_{22}=0.\\
		%\frac{2(q-1)}{q}\binom{q}{k-1}+(-1)^{k-2}\frac{(q-1)(q-2)}{q}\left((-1)^{\lfloor\frac{k-1}{p}\rfloor+1}\binom{\frac{q}{p}-1}{\lfloor\frac{k-1}{p}\rfloor}+(-1)^{\lfloor\frac{k-2}{p}\rfloor}\binom{\frac{q}{p}-1}{\lfloor\frac{k-2}{p}\rfloor} \right),
		%\frac{(q-1)}{q}\binom{q}{k-1}+(-1)^{k-2}\frac{(q-1)^2}{q}\left((-1)^{\lfloor\frac{k-1}{p}\rfloor+1}\binom{\frac{q}{p}-1}{\lfloor\frac{k-1}{p}\rfloor}+(-1)^{\lfloor\frac{k-2}{p}\rfloor}\binom{\frac{q}{p}-1}{\lfloor\frac{k-2}{p}\rfloor} \right),
	\end{cases}$$
\end{theorem}
{\bf Proof}. By Definition \ref{EGRLdefinition}, it's easy to know that $\mathrm{EGRL}_{k,2,0}^{\perp}\left(\mathbb{F}_{q}^{*},\boldsymbol{1},b\right)$ has the generator matrix $\boldsymbol{G}_{ERL}$ given by $(\ref{ERLgenerator})$, thus $\mathrm{EGRL}_{k,2,0}\left(\mathbb{F}_{q}^{*},\boldsymbol{1},b\right)$ is an $[q+2,k]_q$ linear code. And by  Theorem $\ref{ERLdualweight}$, we know that  when $5\leq k\leq q-2$  and $p=2$, or $4\leq k\leq q-1$ and $p\neq 2$, $\mathrm{EGRL}_{k,2,0}^{\perp}\left(\mathbb{F}_{q}^{*},\boldsymbol{1},b\right)$ is AMDS with the parameters $\left[q+2,q+2-k,k\right]$. And then we only need to prove  $$d\left(\mathrm{EGRL}_{k,2,0}\left(\mathbb{F}_{q}^{*},\boldsymbol{1},b\right)\right)=q+2-k.$$ Assume  $$d\left(\mathrm{EGRL}_{k,2,0}\left(\mathbb{F}_{q}^{*},\boldsymbol{1},b\right)\right)\leq q+1-k.$$
Then, for any nonzero codeword $\boldsymbol{c}$ with minimum weight $d$ in $\mathrm{EGRL}_{k,2,0}\left(\mathbb{F}_{q}^{*},\boldsymbol{1},b\right)$, there are at least $k+1$ zero coordinates. Let $\boldsymbol{g}_i$ be the $(i+1)$-th row of $\boldsymbol{G}_{ERL}$, thus we can assume that 
$$
\boldsymbol{c}=\ell_0\boldsymbol{g}_0+\cdots+\ell_{k-1}\boldsymbol{g}_{k-1}=\begin{pmatrix}
	\ell_{0}+\ell_{1}\beta_{1}+\cdots+\ell_{k-1}\beta_{1}^{k-1}\\
	\vdots\\
	\ell_{0}+\ell_{1}\beta_{q-1}+\cdots+\ell_{k-1}\beta_{q-1}^{k-1}\\
	
	\ell_{k-2}a_{11}+\ell_{k-1}a_{21}\\
	\ell_{k-2}a_{12}+\ell_{k-1}a_{22}\\
	\ell_0 b
\end{pmatrix}_{(q+2)\times 1}^T=\left(c_{1},\ldots,c_{q-1},c_{q},c_{q+1},c_{q+2}\right),
$$ 
where $\ell_j\in\mathbb{F}_{q}(j=0,\ldots,k-1)$, and so we have the following five cases.

\textbf{Case 1.} If there exists some codeword $\boldsymbol{c}=(c_{1},\ldots,c_{q-1},c_{q},c_{q+1},c_{q+2})$ with $c_{i_1}=\cdots=c_{i_{k-2}}=c_{q}=c_{q+1}=c_{q+2}=0$ , where $1\leq i_{j}\leq q-1(j=1,\ldots,k-2)$, it means that there exists some $(k-2)$-elements subset $\left\{\beta_{i_1},\ldots, \beta_{i_{k-2}}\right\}$ of $\mathbb{F}_{q}^{*}$ such that
\begin{equation}\label{case11}
\begin{cases}
	\ell_{0}+\ell_{1}\beta_{i}+\cdots+\ell_{k-1}\beta_{i}^{k-1}=0\ ( i=i_1,\ldots,i_{k-2}),\\
		\ell_{k-2}a_{11}+\ell_{k-1}a_{21}=0,\\
		\ell_{k-2}a_{12}+\ell_{k-1}a_{22}=0,\\
		
		\ell_0 b=0.
	\end{cases}
\end{equation}
Note that $b\in\mathbb{F}_{q}^{*}$, it means $\ell _0=0$.  And by $\det \begin{pmatrix}
	a_{11}&a_{21}\\
	a_{12}&a_{22}
\end{pmatrix}=a_{11}a_{22}-
a_{21}a_{12}=\det(\boldsymbol{M}_{2\times 2}^{T})\neq 0,$ the homegeneous systen of the equations $\begin{cases}
\ell_{k-2}a_{11}+\ell_{k-1}a_{21}=0,\\
\ell_{k-2}a_{12}+\ell_{k-1}a_{22}=0,
\end{cases}$ has only zero solution, i.e., $\ell_{k-2}=\ell_{k-1}=0.$ Furthermore, by (\ref{case11}) we know that for any $i\in\{i_1,\ldots,i_{k-2}\},$
$$\ell_{1}\beta_{i}+\cdots+\ell_{k-3}\beta_{i}^{k-3}=0,$$
i.e., the polynomial $f_{1}(x)=\ell_{1}x+\cdots+\ell_{k-3}x^{k-3}$ has $k-2$ roots over $\mathbb{F}_{q}$, which is a contradiction with $\deg(f(x))=k-3$.

\textbf{Case 2.} If there exists some codeword $\boldsymbol{c}=(c_{1},\ldots,c_{q-1},c_{q},c_{q+1},c_{q+2})$ with $c_{i_1}=\cdots=c_{i_{k-1}}=c_{q}=c_{q+1}=0$ , where $1\leq i_{j}\leq q-1(j=1,\ldots,k-1)$, it means that there exists some $(k-1)$-elements subset $\left\{\beta_{i_1},\ldots, \beta_{i_{k-1}}\right\}$ of $\mathbb{F}_{q}^{*}$ such that 
\begin{equation}\label{case22}
	\begin{cases}
				\ell_{0}+\ell_{1}\beta_{i}+\cdots+\ell_{k-1}\beta_{i}^{k-1}=0\ ( i=i_1,\ldots,i_{k-1}),\\
				\ell_{k-2}a_{11}+\ell_{k-1}a_{21}=0,\\
		\ell_{k-2}a_{12}+\ell_{k-1}a_{22}=0.
	\end{cases}
\end{equation}
Note that $\det \begin{pmatrix}
	a_{11}&a_{21}\\
	a_{12}&a_{22}
\end{pmatrix}=a_{11}a_{22}-
a_{21}a_{12}=\det(\boldsymbol{M}_{2\times 2}^{T})\neq 0,$ it means that the homegeneous systen of the equations  $\begin{cases}
	\ell_{k-2}a_{11}+\ell_{k-1}a_{21}=0,\\
	\ell_{k-2}a_{12}+\ell_{k-1}a_{22}=0,
\end{cases}$ has only zero solution, i.e., $\ell_{k-2}=\ell_{k-1}=0.$ Furthermore, by (\ref{case22}) we know that for any $i\in\{i_1,\ldots,i_{k-1}\},$
$$\ell_{0}+\ell_{1}\beta_{i}+\cdots+\ell_{k-3}\beta_{i}^{k-3}=0,$$
i.e., the polynomial $f_{2}(x)=\ell_{0}+\ell_{1}x+\cdots+\ell_{k-3}x^{k-3}$ has $k-1$ roots over $\mathbb{F}_{q}$, which is a contradiction with $\deg(f_{2}(x))=k-3$.

\textbf{Case 3.} If there exists some codeword $\boldsymbol{c}=(c_{1},\ldots,c_{q-1},c_{q},c_{q+1},c_{q+2})$ with $c_{i_1}=\cdots=c_{i_{k-1}}=c_{q+2}=0$ and one of $c_{q}$ and $c_{q+1}$ is zero, where $1\leq i_{j}\leq q-1(j=1,\ldots,k-1)$, it means that there exists some $(k-1)$-elements subset $\left\{\beta_{i_1},\ldots, \beta_{i_{k-1}}\right\}$ of $\mathbb{F}_{q}^{*}$ such that
\begin{equation}\label{case31}
	\begin{cases}
		\ell_{0}+\ell_{1}\beta_{i}+\cdots+\ell_{k-1}\beta_{i}^{k-1}=0\ ( i=i_1,\ldots,i_{k-1}),\\
		\ell_{k-2}a_{11}+\ell_{k-1}a_{21}=0,\\
		\ell_{0} b=0;
	\end{cases}
\end{equation}
or 
\begin{equation}\label{case32}
	\begin{cases}
		\ell_{0}+\ell_{1}\beta_{i}+\cdots+\ell_{k-1}\beta_{i}^{k-1}=0\ ( i=i_1,\ldots,i_{k-1}),\\
		\ell_{k-2}a_{12}+\ell_{k-1}a_{22}=0,\\
		\ell_{0} b=0.
	\end{cases}
\end{equation}
Note that $b\in\mathbb{F}_{q}^{*},$ it means $\ell_{0}=0.$ Furthermore, by (\ref{case31})-(\ref{case32}), we know that for any $i\in\{i_1,\ldots,i_{k-1}\},$
$$\ell_{1}\beta_{i}+\cdots+\ell_{k-1}\beta_{i}^{k-1}=0,$$
i.e., 
$$\beta_{i}\left(\ell_{1}+\ell_{2}\beta_{i}^{2}+\cdots+\ell_{k-1}\beta_{i}^{k-2}\right)=0.$$
Note that $\beta_{i}\in\mathbb{F}_{q}^{*}$, it means that for any $i\in\{i_1,\ldots,i_{k-1}\},$
$$\ell_{1}+\ell_{2}\beta_{i}^{2}+\cdots+\ell_{k-1}\beta_{i}^{k-2}=0,$$
the polynomial $f_{3}(x)=\ell_{1}+\ell_{2}x+\cdots+\ell_{k-1}x^{k-2}$ has $k-1$ roots over $\mathbb{F}_{q}$, which is a contradiction with $\deg(f_{3}(x))=k-2$.

\textbf{Case 4.} If there exists some codeword $\boldsymbol{c}=(c_{1},\ldots,c_{q-1},c_{q},c_{q+1},c_{q+2})$ such that $c_{i_1}=\cdots=c_{i_{k}}=0$ and one of $c_{q}$, $c_{q+1}$ and $c_{q+2}$ is zero, where $1\leq i_{j}\leq q-1(j=1,\ldots,k)$, it means that there exists some  $k$-elements subset $\left\{\beta_{i_1},\ldots, \beta_{i_{k}}\right\}$ of $\mathbb{F}_{q}^{*}$ such that
\begin{equation}\label{case41}
	\begin{cases}
		\ell_{0}+b_{1}\beta_{i}+\cdots+\ell_{k-1}\beta_{i}^{k-1}=0(i=i_1,\ldots,i_{k}), \\
		\ell_{k-2}a_{11}+\ell_{k-1}a_{21}=0;
	\end{cases}
\end{equation}
or
\begin{equation}\label{case42}
	\begin{cases}
		\ell_{0}+\ell_{1}\beta_{i}+\cdots+\ell_{k-1}\beta_{i}^{k-1}=0(i=i_1,\ldots,i_{k}), \\
		\ell_{k-2}a_{12}+\ell_{k-1}a_{22}=0;
	\end{cases}
\end{equation}
or
\begin{equation}\label{case43}
	\begin{cases}
		\ell_{0}+\ell_{1}\beta_{i}+\cdots+\ell_{k-1}\beta_{i}^{k-1}=0(i=i_1,\ldots,i_{k}), \\
		\ell_{0}b=0.
	\end{cases}
\end{equation}
Furthermore, by $(\ref{case41})$-$(\ref{case43})$, we know that the polynomial $f_{4}(x)=\ell_{0}+_{1}x+\cdots+b_{k-1}x^{k-1}$ has $k$ roots over $\mathbb{F}_{q}$, which is a contradiction with $\deg(f_{4}(x))=k-1$.

\textbf{Case 5.} If there exists some codeword $\boldsymbol{c}=(c_{1},\ldots,c_{q-1},c_{q},c_{q+1},c_{q+2})$ with $c_{i_1}=\cdots=c_{i_{k+1}}=0$, where $1\leq i_{j}\leq q-1(j=1,\ldots,k+1)$, it means that there exists some $(k+1)$-elements subset $\left\{\beta_{i_1},\ldots, \beta_{i_{k+1}}\right\}$ of $\mathbb{F}_{q}^{*}$ such that
$$\ell_{0}+\ell_{1}\beta_{i}+\cdots+\ell_{k-1}\beta_{i}^{k-1}=0\ (i=i_1,\ldots,i_{k+1}),$$
it means that the polynomial $f_{5}(x)=\ell_{0}+\ell_{1}x+\cdots+\ell_{k-1}x^{k-1}$ has $k+1$ roots over $\mathbb{F}_{q}$, 
which is a contradiction with $\deg(f_{5}(x))=k-1$.

From the above discussions, we have $d\left(\mathrm{EGRL}_{k,2,0}\left(\mathbb{F}_{q}^{*},\boldsymbol{1},b\right)\right)\geq q+2-k$. Note that by the Singleton bound, we have $d\left(\mathrm{EGRL}_{k,2,0}\left(\mathbb{F}_{q}^{*},\boldsymbol{1},b\right)\right) \leq q+3-k $, then
$$q+2-k\leq d\left(\mathrm{EGRL}_{k,2,0}\left(\mathbb{F}_{q}^{*},\boldsymbol{1},b\right)\right) \leq q+3 - k.$$
If $d\left(\mathrm{EGRL}_{k,2,0}\left(\mathbb{F}_{q}^{*},\boldsymbol{1},b\right)\right)=q+3-k$, then $\mathrm{EGRL}_{k,2,0}\left(\mathbb{F}_{q}^{*},\boldsymbol{1},b\right)$ is MDS with the parameters $[q+2,k,q+3-k]_q$, and so by Lemma \ref{MDSdefinition},  $\mathrm{EGRL}_{k,2,0}^{\perp}\left(\mathbb{F}_{q}^{*},\boldsymbol{1},b\right)$ is also  MDS, which contradicts with Theorem $\ref{ERLdualweight}$.
It means that $d\left(\mathrm{EGRL}_{k,2,0}\left(\mathbb{F}_{q}^{*},\boldsymbol{1},b\right)\right)=q+2-k$, i.e.,  $\mathrm{EGRL}_{k,2,0}\left(\mathbb{F}_{q}^{*},\boldsymbol{1},b\right)$ is NMDS with the parameters $[q+2,k,q+2-k]_q$. Furthermore by Lemma  \ref{NMDSweight}, $A_{q+2-k}=A_{k}^{\perp}$.

This completes the proof of Theorem $\ref{ERL1}$.

$\hfill\Box$

By Theorem $\ref{ERL1}$, Lemma \ref{NMDSweight} and directly computing, one can obtain the following Corollary \ref{EGRLcorollary} immediately. 

\begin{corollary}\label{EGRLcorollary}
If $q=3^m$, $k=5$, $b\in\mathbb{F}_{q}^{*}$ and $\boldsymbol{M}_{2\times 2}= \begin{pmatrix}
	a_{11}&a_{12}\\
	a_{21}&a_{22}
\end{pmatrix}$ with $a_{11}a_{12}a_{21}a_{22}\neq 0$, then
\iffalse $$\boldsymbol{G}_{ERL}=
	\begin{pmatrix}
		1 &          \cdots & 1& 0 & 0              & 2   \\
		\beta_1   &  \cdots & \beta_{q-1}   & 0 & 0 & 0 \\
		\beta_1^2 &  \cdots & \beta_{q-1}^2 & 0 & 0 & 0 \\
		\beta_1^3 &  \cdots & \beta_{q-1}^3  & 1 & 1& 0 \\
		\beta_1^4 &  \cdots & \beta_{q-1}^4  & 2 & 1& 0 
	\end{pmatrix}.
	$$ 
then\fi $\mathrm{EGRL}_{k,2,0}\left(\mathbb{F}_{q}^{*},\boldsymbol{1},b\right)$ is NMDS with the parameters $[3^{m}+2,5,3^{m}-3]$, and
\iffalse$$\begin{aligned}
		A_{q-3}=&A_5^\perp\\
		=&(q-1)\# N(4,2,\mathbb{F}_{q}^{*})+(q-1)\# N(4,1,\mathbb{F}_{q}^{*})+(q-1)\# N(3,2,\mathbb{F}_{q}^{*})+(q-1)\# N(3,1,\mathbb{F}_{q}^{*})\\
		=&\frac{(q-1)^2(q-2)(q-3)}{12}.
	\end{aligned}$$ By Lemma \ref{NMDSweight},\fi the weight enumerator of $\mathrm{EGRL}_{k,2,0}\left(\mathbb{F}_{q}^{*},\boldsymbol{1},b\right)$ is	
	$$
	\begin{aligned}
		A(x)=&1+A_{q-3}x^{q-3}+A_{q-2}x^{q-2}+A_{q-1}x^{q-1}+A_{q}x^{q}+A_{q+1}x^{q+1}+A_{q+2}x^{q+2}\\
		=&1+\frac{(q-1)^2(q-2)(q-3)}{12}x^{q-3}+\frac{(q-1)^2(q^3-7q^2+52q-60)}{24} x^{q-2}\\
		&+\frac{(q-1)(7q^3-24q^2+59q-30)}{6}x^{q-1}+\frac{(q-1)(3q^4-4q^3+63q^2-98q+72)}{12}x^{q}\\
		&+\frac{(q-1)^2(4q^3+17q^2-17q+30)}{12}x^{q+1}+\frac{(q-1)(9q^4-16q^3+15q^2-20q+12)}{24}x^{q+2}.
	\end{aligned}
	$$

\end{corollary} 

Next, we give an example for Corollary \ref{EGRLcorollary}.
\begin{example}\label{ERL1example} 
Let $m=b=2$, $\mathbb{F}_{3^2}^{*}=\langle\omega\rangle$ and $\beta_{i}=\omega^{i-1}(i=1,\ldots,8)$, then
$$\begin{pmatrix}
1 &1&1&1&1&1 & 1&1& 0 & 0& 2  \\
1   &\omega&\omega^{2}&\omega^{3}&2&\omega^{5}&\omega^{6}&\omega^{7}& 0& 0 & 0 \\
1 &\omega^{2}&2&\omega^{6}&1&\omega^{2}&2&\omega^{6} & 0& 0 & 0 \\
1 &\omega^{3}&\omega^{6}&\omega&2&\omega^{7}&\omega^{2}&\omega^{5} & 1 & 1& 0 \\
	1 &2&1&2&1&2&1&2& 2 & 1& 0 \\
\end{pmatrix}.
$$
is the generator matrix of $\mathrm{EGRL}_{k,2,0}\left(\mathbb{F}_{q}^{*},\boldsymbol{1},b\right)$. 
Based on the Magma programe, $\mathrm{EGRL}_{k,2,0}\left(\mathbb{F}_{q}^{*},\boldsymbol{1},b\right)$ is a $\mathbb{F}_{3^2}$-linear code with the parameters $[11,5,6]_{3^2}$ and its weight enumerator is
$$A(x)=1+224x^6+1520x^7+4880x^8+14040x^9+22240x^{10}+16144x^{11},$$
which is consistent with Corollary \ref{EGRLcorollary}.
\end{example}
\section{Conclusions}
In this paper, we define a class of the extended code of generalized Roth-Lempel codes $\mathrm{EGRL}_{k,\ell,t}(\boldsymbol{\alpha}, \boldsymbol{v},b)$, discuss some coding properties of a class of EGRL codes $\mathrm{EGRL}_{k,2,0}(\boldsymbol{\alpha}, \boldsymbol{v},b)$, construct a class of NMDS codes and obtain the following main results.
\begin{itemize}
\item A parity-check matrix of $\mathrm{EGRL}_{k,2,0}(\boldsymbol{\alpha}, \boldsymbol{v},b)$ (Theorem \ref{EGRLparitycheckmatrix});
\item An equivalent condition for $\mathrm{EGRL}_{k,\ell,t}(\boldsymbol{\alpha}, \boldsymbol{v},b)$ to be MDS (Theorem \ref{ERLMDS});
\item An equivalent condition for $\mathrm{EGRL}_{k,\ell,t}^{\perp}(\boldsymbol{\alpha}, \boldsymbol{v},b)$ to be AMDS (Theorem \ref{ERLAMDS});
\item A class of NMDS EGRL codes $\mathrm{EGRL}_{k,2,0}\left(\mathbb{F}_{q}^{*},\boldsymbol{1},b\right)$  is given and their weight distributions are completely determined (Theorem \ref{ERL1}). Especially, when $\begin{pmatrix}
	a_{11}&a_{12}\\
	a_{21}&a_{22}
\end{pmatrix}=\begin{pmatrix}
	0&1\\
	1&0
\end{pmatrix}$ and $b=1$, Theorem \ref{ERL1} is just Theorem 2 in the reference \cite{A4}. 
\end{itemize}
 
Additionally, from the proofs of Theorems \ref{ERLMDS} and \ref{ERLAMDS}, we know  that the establishment of the equivalence condition for \(\mathrm{EGRL}_{k,2,0}(\boldsymbol{\alpha}, \boldsymbol{v},b)\) or its dual to be MDS or AMDS only depends on the non-zeroity of $b$, and not depends on the specific value of $b$. Therefore, without loss of generality, we can take \(b=1\) later. 

\end{document}